\documentclass[twocolumn, times]{aastex631} 

\shorttitle{$H_0$ from lensed SNe}
\shortauthors{Birrer et al.}
\graphicspath{{./}{figures/}}

\newcommand{\Hunit}{km s$^{-1}{\rm Mpc}^{-1}$}
\newcommand{\notebooklink}{https://github.com/sibirrer/glSNe/tree/main/Notebooks/notebook.ipynb}

\usepackage{amsmath}    
\usepackage{bbold}  
\usepackage{fontawesome}  

\begin{document}

\title{The Hubble constant from strongly lensed supernovae with standardizable magnifications}

\author[0000-0003-3195-5507]{Simon Birrer}
\affiliation{Kavli Institute for Particle Astrophysics and Cosmology and Department of Physics, Stanford University, Stanford, CA 94305, USA}
\affiliation{SLAC National Accelerator Laboratory, Menlo Park, CA, 94025}
\correspondingauthor{Simon Birrer}
\email{sibirrer@stanford.edu}

\author[0000-0002-2376-6979]{Suhail Dhawan}
\affiliation{Kavli Institute of Cosmology / Institute of Astronomy, University of Cambridge Madingley Road, Cambridge CB3 0HA, UK}

\author[0000-0002-5558-888X]{Anowar J. Shajib}
\affiliation{Department of Astronomy \& Astrophysics, University of Chicago, Chicago, IL 60637, USA}



\begin{abstract}

The dominant uncertainty in the current measurement of the Hubble constant ($H_0$) with strong gravitational lensing time delays is attributed to uncertainties in the mass profiles of the main deflector galaxies. Strongly lensed supernovae (glSNe) can provide, in addition to measurable time delays, lensing magnification constraints when knowledge about the unlensed apparent brightness of the explosion is imposed.
We present a hierarchical Bayesian framework to combine a dataset of SNe that are not strongly lensed and a dataset of strongly lensed SNe with measured time delays. We jointly constrain (i) $H_0$ using the time delays as an absolute distance indicator, (ii) the lens model profiles using the magnification ratio of lensed and unlensed fluxes on the population level and (iii) the unlensed apparent magnitude distribution of the SNe population and the redshift-luminosity relation of the relative expansion history of the Universe. 
We apply our joint inference framework on a future expected data set of glSNe, and forecast that a sample of 144 glSNe of Type~Ia with well measured time series and imaging data will measure $H_0$ to 1.5\%.
We discuss strategies to mitigate systematics associated with using absolute flux measurements of glSNe to constrain the mass density profiles. Using the magnification of SNe images is a promising and complementary alternative to using stellar kinematics.
Future surveys, such as the Rubin and \textit{Roman} observatories, will be able to discover the necessary number of glSNe, and with additional follow-up observations this methodology will provide precise constraints on mass profiles and $H_0$. \href{https://github.com/sibirrer/glSNe}{\faGithub}

\end{abstract}



\section{Introduction}
The current expansion rate of the Universe, the Hubble Constant $H_0$, anchors the scale and the age of the Universe.
There is an ongoing debate about the precise value of $H_0$, where some local distance ladder measurements based on calibration using Cepheids \citep[e.g.,][]{Riess:2021} are in significant statistical disagreement with measurements extrapolated from the cosmic microwave background (CMB) \citep[e.g.,][]{Planck:2020cosmo_param, Aiola:2020}. Another distance ladder analysis based on calibration using the tip of the red giant branch (TRGB) stars results in a consistent measurement with the CMB \citep{Freedman:2020, Freedman:2021}. This discrepancy either indicates unaccounted systematics in one or multiple measurements \citep[e.g.,][]{Efstathiou:2020, Mortsell:2021}, or new physics beyond the standard model of cosmology.
Multiple independent and precise measurements of $H_0$ are essential in providing a definite resolution to the current tension.

Relative time delays between multiple gravitationally lensed images provide a one-step distance anchor of the Universe, and thus $H_0$. This probe is independent of the local distance ladder and the sound-horizon-physics anchors of the CMB and large-scale structure probes. The method, known as the time-delay cosmography, has been proposed more than half a century ago to utilize the transient nature of supernovae (SNe) for measuring the time delays \citep[][]{Refsdal:1964}.
Time-delay cosmography was first applied by measuring the time delays of multiply lensed quasars with multi-season monitoring campaigns \citep[e.g.,][]{Kundic:1997, Schechter:1997, Fassnacht:2002, Tewes:2013, Courbin:2018, Millon:2020cosmograil}.
The discovery of numerous lensed quasar systems, follow-up monitoring, high-resolution imaging and precise spectroscopic observations have lead to a precise measurement of $H_0$ using seven multiply lensed quasars \citep{Wong:2020, Shajib:2020strides, Millon:2020}. These measurements assumed particular forms of the mass density profiles of the deflector galaxies. The mass-sheet degeneracy \citep[see][hereafter, MSD]{Falco:1985, SchneiderSluse:2013}, an inherent transform leaving the lensing observables invariant while changing the time-delay prediction, poses limits in the precision of $H_0$ measurements in the absence of additional data. \cite{Birrer:2020} introduced an additional degree of freedom to the mass density profiles to avoid constraining the lens model based on the specific form of the mass profiles previously chosen. \cite{Birrer:2020} constrained the MSD solely by stellar kinematics observations of the deflector galaxy hierarchically on the deflector population level mitigating covariances among the assumptions of individual lenses. For the achieved 5\% precision measurement of $H_0$, \cite{Birrer:2020} combined the 7 TDCOSMO lenses with 33 galaxy--galaxy lenses from the Sloan Lens ACS (SLACS) survey \citep{Bolton:2008, Shajib:2021}. The interpretation of the kinematics measurements are impacted by the mass-anisotropy degeneracy \citep{BinneyMamon:1982, Dejonghe:1992} and mitigating this degeneracy requires assumption on the stellar anisotropy distribution or spatially resolved kinematics measurements \citep[e.e.,][]{Cappellari:2008, Barnabe:2011, Yildirim:2020}. A forecast for future constraints using kinematics observations in breaking the MSD within the assumptions of the \cite{Birrer:2020} analysis is provided by \cite{BirrerTreu:2021}.

An alternative to lensed quasars, as in fact anticipated in the original work by \cite{Refsdal:1964}, are multiply-resolved gravitationally lensed supernovae (glSNe).
glSNe are exquisite laboratories for fundamental physics, as well as astrophysical properties of the host and lens galaxies  \citep[see][for a review of strong lensing of SNe and other explosive transients]{2019RPPh...82l6901O}. We refer to
\citet{Goobar:2002}, for example, for early explorations of cosmological parameter forecasts with hundreds of glSNe.

Although strongly lensed galaxies and quasars (QSOs) are more common than currently discovered occurrences of glSNe, glSNe have notable advantages, particularly if they are of Type Ia (SNe~Ia). The luminosities of SNe~Ia have a small dispersion after correcting for the relations with the lightcurve shape, observed color, and properties of their host galaxies, making them a ``standardizable candle" \citep[e.g., see][]{Phillips:1993, Guy:2007js, Scolnic:2018}.
Knowledge about the apparent magnitude, at the source redshift of the SNe, in the absence of any lensing effect allows us to directly measure the lensing magnification factor at the locations relevant to predict the time delays, breaking the MSD \citep[see also e.g.,][]{Kolatt:1998, Oguri:2003, Foxley-Marrable:2018}. Thus, breaking the MSD does not require an 
\emph{a priori} knowledge of the SN~Ia absolute magnitude, and thus keeping the inference from time-delay cosmography independent from the local distance ladder calibration. Additionally, SNe~Ia have a well-studied family of light curves with a well-defined maximum at $
\sim$18 days from explosion \citep[e.g.][]{Yao2019,Miller:2020} and hence, can be used for an accurate measurement of time delays, with significantly fewer follow-up observations than quasars. Recent simulations of lensed SNe~Ia also find that the time-delay measurement is not impacted significantly from microlensing \citep{Goldstein:2018, Huber:2021}.  Moreover, since SNe fade away, we can obtain post-explosion imaging to validate the lens model \citep[see e.g.,][]{Ding:2021}. 

glSNe~Ia are also complementary to lensed QSOs in the strategy for discovering these system. Owing to their small luminosity scatter, glSNe~Ia can be discovered due to the lensing magnification increasing their brightness. This would not need  highly spatially resolved observations, as is the case for lensed QSOs, important for testing potential biases from selecting high angular separation events. This was demonstrated in the discovery of the first resolved strongly lensed SN~Ia, iPTF16geu \citep{Goobar:2017}. At $z=0.409$, the SNe was found to be 30 standard deviations too bright compared with the SNe~Ia population, prompting space-based and laser guided star-adaptive optics (LGS-AO) follow-up. While iPTF16geu had a short time delay of $\sim 1$ day \citep{Goobar:2017, More:2017}, hence, not ideal for measuring $H_0$, the system could uniquely be used for a direct inference of the lensing magnification \citep{Dhawan:2020}. Simulations of wide-field surveys like the Zwicky Transient Factory (ZTF) suggest a median time delay of $\sim 1-5$ days \citep{Goldstein:2019, Wojtak:2019} based on a magnification discovery channel. Upcoming deeper surveys, such as the Vera Rubin Observatory Legacy Survey of Space and Time (LSST), are expected to discover glSNe based on image multiplicity at fainter magnitudes, shifting the median time delay to $\sim 10$ days \citep{Wojtak:2019}, making glSNe~Ia compelling probes of $H_0$. glSNe also offer a unique opportunity to obtain time-delays from resolved spectroscopy, a method which requires very few epochs of observations \citep{Johansson:2021, Bayer:2021}.

With the advent of transient astrophysics and the anticipated discovery of more glSNe from current and future time-domain facilities, glSNe can play a major and complementary role in time-delay cosmography and beyond. In particular, the complementarity in constraining the MSD with magnification measurements in addition to stellar kinematics measurements allows one to rigorously check for systematics inherent in either of the two approaches as well as gain further statistical precision in the most limiting domain of time-delay cosmography to date.

Additionally, strong gravitational lensing systems are powerful probe of elliptical galaxy properties and evolution \citep[e.g.,][]{Treu:2002, Auger:2010, Shajib:2021}. Non-imaging data -- such as the time delays, the stellar kinematics, or the image magnifications -- provide additional constraints on the gravitational potential, or equivalently the mass distribution. In such studies, the adopted values of cosmological parameters can indeed have significant physical outcomes. For example, \citet{Blum:2020} demonstrated that adopting the CMB-based $H_0$ value for the 7 TDCOSMO systems leads to galaxy mass distributions with a cored component in the dark matter profile. However, \citet{Shajib:2021} combined only the stellar kinematics with the lens imaging data (without any time-delay measurement) to find that the deviation from the power-law profile in elliptical galaxies can also be caused by a higher normalization in the dark matter profile instead of having a cored component. The hierarchical analysis of \citet{Birrer:2020} simultaneously constrained both the mass distribution in galaxies and $H_0$ for the first time from the combination of stellar kinematics and lensing information. glSN~Ia will similarly provide simultaneous constraints on the galaxy mass distribution and cosmological parameters. Furthermore, a sufficiently large glSN~Ia sample spanning a wide redshift range can provide direct insights into the evolution of massive elliptical galaxies.

In this paper, we aim to exploit the uniform behaviour of the population of SNe~Ia to reduce uncertainties in the lens modeling arising from the MSD. We extend the hierarchical inference framework by \cite{Birrer:2020} and incorporate SNe~Ia apparent magnitudes, for both lensed and global unlensed populations, on the likelihood level in the cosmographic inferences. We perform forecasts at the same level of complexity as presented by \cite{BirrerTreu:2021}, now replacing the kinematics observables with SNe~Ia brightness measurements, for different scenarios and highlight the key ingredients required to achieve an $H_0$ measurement with precision below $2\%$.

The paper is structured as follows. Section~\ref{sec:general_tdc} provides a general review on the key concepts of time-delay cosmography, with a special focus on the MSD and approaches to constrain it.
Section~\ref{sec:method} defines the methodology, model parameterization, likelihood, and sampling approach used for the forecasts.
Section~\ref{sec:forecasts} presents different forecast scenarios in regard to $H_0$ inferences. We discuss the key components and implications of this work in Section~\ref{sec:discussion} and conclude in Section~\ref{sec:conclusion}.

The formalism and inference schemes presented in this work are implemented in the open-source software \textsc{hierArc}\footnote{\url{https://github.com/sibirrer/hierArc}} and the scripts to reproduce the presented work is publicly available\footnote{\url{https://github.com/sibirrer/glSNe}}. Lensing calculations are performed with \textsc{lenstronomy}\footnote{\url{https://github.com/sibirrer/lenstronomy}} \citep{lenstronomy1, lenstronomy2}.

\section{Time-delay cosmography with strongly lensed SNe}\label{sec:general_tdc}

In this section, we review the principles of time-delay cosmography for lensing and time delays (Section \ref{sec:td_cosmo}). We then emphasize how an MSD affects the observables and thus the inference of cosmographic quantities, and specifically discuss the ability of glSNe in breaking the MSD with absolute lensing magnifications (Section \ref{sec:mst_glsne}).

\subsection{Cosmography with strong lenses} \label{sec:td_cosmo}

The phenomena of gravitational lensing can be described by the lens equation, which maps the source plane coordinate $\boldsymbol{\beta}$ to the image plane $\boldsymbol{\theta}$ as
\begin{equation} \label{eqn:lens_equation}
  \boldsymbol{\beta} = \boldsymbol{\theta} - \boldsymbol{\alpha}(\boldsymbol{\theta}),
\end{equation}
where $\boldsymbol{\alpha}$ is the angular shift on the sky between the original unlensed position and the lensed observed position of an object.

For a single deflector plane, the lens equation can be expressed in terms of the physical deflection angle $\hat{\boldsymbol{\alpha}}$ as
\begin{equation} \label{eqn:lens_equation_single_plane}
  \boldsymbol{\beta} = \boldsymbol{\theta} - \frac{D_{\rm s}}{D_{\rm ds}}\hat{\boldsymbol{\alpha}}(\boldsymbol{\theta}),
\end{equation}
where $D_{\rm s}$ and $D_{\rm ds}$ are the angular diameter distances from the observer to the source and from the deflector to the source, respectively.
In the single lens plane regime, we can introduce the lensing potential $\psi$ such that
\begin{equation}
    \boldsymbol{\alpha}(\boldsymbol{\theta}) = \nabla \psi(\boldsymbol{\theta}),
\end{equation}
and the lensing convergence as
\begin{equation}
    \kappa(\boldsymbol{\theta}) =  \frac{1}{2}\nabla^2 \psi(\boldsymbol{\theta}).
\end{equation}
The relative arrival time $\Delta t_{\rm AB}$ between two images $\boldsymbol{\theta}_{\rm A}$ and $\boldsymbol{\theta}_{\rm B}$ originated from the same source is
\begin{equation}\label{eqn:time_delay}
    \Delta t_{\rm AB} = \frac{D_{\Delta t}}{c} \left[\tau(\boldsymbol{\theta}_{\rm A}, \boldsymbol{\beta}) - \tau(\boldsymbol{\theta}_{\rm B}, \boldsymbol{\beta}) \right] = \frac{D_{\Delta t}}{c} \Delta\tau_{\rm AB},
\end{equation}
where $c$ is the speed of light,
\begin{equation}\label{eqn:fermat_potential}
    \tau(\boldsymbol{\theta}, \boldsymbol{\beta}) = \left[ \frac{\left(\boldsymbol{\theta} - \boldsymbol{\beta} \right)^2}{2} - \psi(\boldsymbol{\theta})\right]
\end{equation}
is the Fermat potential \citep{Schneider:1985, Blandford:1986}, and
\begin{equation} \label{eqn:ddt_definition}
    D_{\Delta t} \equiv \left(1 + z_{\rm d}\right) \frac{D_{\rm d}D_{\rm s}}{D_{\rm ds}},
\end{equation}
is the time-delay distance \citep{Refsdal:1964, Schneider:1992, Suyu:2010}; $D_{\rm d}$ is the angular diameter distance from the observer to the deflector. In the last line of Equation~\ref{eqn:time_delay} we chose the notation $\Delta\tau_{\rm AB}$ to describe the relative Fermat potential between two images.

Constraints on the Fermat potential difference $\Delta \tau_{\rm AB}$ and a measured time delay $\Delta t_{\rm AB} $ allow us to constrain the time-delay distance $D_{\Delta t}$. This absolute physical distance anchors the scale in the Universe within the redshifts involved in the lensing configuration.
The Hubble constant is inversely proportional to the absolute scales of the Universe and thus scales with $D_{\Delta t}$ as
\begin{equation} \label{eqn:H0_ddt}
	H_0 \propto D_{\Delta t}^{-1},
\end{equation}
mildly dependent on the relative expansion history from current time ($z=0$) to the redshifts of the deflector and the source.

\subsection{The MST and the ability of lensing magnifications in breaking it} \label{sec:mst_glsne}

\subsubsection{MST impact on time delays and imaging data}
The mass-sheet transform (MST) is a multiplicative transform of the lens equation (Eqn. \ref{eqn:lens_equation}) given by

\begin{equation} \label{eqn:lens_equation_MST}
  \lambda \boldsymbol{\beta} = \boldsymbol{\theta} - \lambda \boldsymbol{\alpha}(\boldsymbol{\theta}) - (1 - \lambda) \boldsymbol{\theta},
\end{equation}
which preserves image positions (and any higher order relative differentials of the lens equation) under a linear source displacement $\boldsymbol{\beta} \rightarrow \lambda\boldsymbol{\beta}$ \citep{Falco:1985}.
The term $(1 - \lambda) \boldsymbol{\theta}$ in Equation \ref{eqn:lens_equation_MST} above describes an infinite sheet of convergence (or mass), and hence the name mass-sheet transform. Only observables related to the unlensed apparent source size, to the unlensed apparent brightness, or to the lensing potential are able to break this degeneracy.

The convergence field transforms according to
\begin{equation}\label{eqn:mst}
    \kappa_{\lambda}(\theta) = \lambda \kappa(\theta) + \left( 1 - \lambda\right).
\end{equation}
Thus, the same relative lensing observables can result if the mass profile is scaled by the factor $\lambda$ with the addition of a sheet of convergence (or mass) of $\kappa(\boldsymbol{\theta}) = (1-\lambda)$.

The different observables described in Section \ref{sec:td_cosmo} relevant for time-delay cosmography transform by an MST term $\lambda$ as follows:
the image positions remain invariant
\begin{equation}
    \boldsymbol{\theta}_{\lambda} = \boldsymbol{\theta};
\end{equation}
the source position scales with $\lambda$ as
\begin{equation}
    \boldsymbol{\beta}_{\lambda} = \lambda \boldsymbol{\beta};
\end{equation}
the Fermat potential scales with $\lambda$ as
\begin{equation}\label{eqn:fermat_mst}
    \Delta \tau_{\rm AB , \lambda} =  \lambda \Delta \tau_{\rm AB},
\end{equation}
and so does the time delay as
\begin{equation}\label{eqn:time_delay_mst}
    \Delta t_{\rm AB , \lambda} =  \lambda \Delta t_{\rm AB}.
\end{equation}

When transforming a deflector profile with an MST, the inference of the time-delay distance (Eqn.~\ref{eqn:ddt_definition}) from a measured time delay and inferred Fermat potential transforms as
\begin{equation} \label{eqn:ddt_mst}
    D_{\Delta t , \lambda} = \lambda^{-1}D_{\Delta t}.
\end{equation}
Thus, the Hubble constant, when inferred from the time-delay distance $D_{\Delta t}$, transforms as (from Eqn.~\ref{eqn:H0_ddt})
\begin{equation} \label{eqn:h0_mst}
H_{0 , \lambda} =  \lambda H_0.
\end{equation}

Achieving precise and accurate constraints on the radial density profile required to measure $H_0$ necessitates external data and puts high demand on the precision and accuracy of those measurements and priors.
We refer the reader to Section~2 of \cite{Birrer:2020} for a discussion on interpretations of an MST in regard to a parameterized profile and physical limits of it.

There are two promising observables that have the ability to break the MST independent of the time delays: the stellar velocity dispersion measurements of the deflector galaxy and the absolute magnification measurement from knowledge of the apparent unlensed brightness of a source component.

For the remainder of this paper, we chose the convention of $\lambda$ to be the mapping from a model prediction ignoring MST effects to the target prediction of the correct answer.

\subsubsection{Stellar velocity dispersion}
The stellar velocity dispersion of the main deflector galaxy is directly sensitive to the deflector potential. Joint lensing and kinematics measurements have been used to constrain the mass profiles of massive elliptical galaxies \citep{Shajib:2021} and is the sole constraining anchor on the MST in the $H_0$ measurement by \cite{Birrer:2020}. The observed stellar velocity dispersion $\sigma_{\rm v}$ scales with an MST as
\begin{equation} \label{eqn:kinematics_mst}
    \sigma_{\rm v , \lambda} = \sqrt{\lambda} \sigma_{\rm v}.
\end{equation}

A fractional uncertainty in the velocity dispersion measurement $\sigma_{\rm v, obs}$,  or model prediction $\sigma_{\rm v, model}$ propagates to a fractional uncertainty in the MST as
\begin{equation}\label{eqn:v_disp_mst}
    \frac{\delta \lambda}{\lambda} = 2 \left[\frac{\delta \sigma_{\rm v, obs}}{\sigma_{\rm v, obs}} - \frac{\delta \sigma_{\rm v, model}}{\sigma_{\rm v, model}} \right],
\end{equation}
where we identified the target truth (measured) velocity dispersion with $\sigma_{\rm v , \lambda} = \sigma_{\rm v, obs}$ and the model without the MST correction with $\sigma_{\rm v} = \sigma_{\rm v, model}$ of Equation~\ref{eqn:kinematics_mst}.
Thus, an achievable 5\% uncertainty in the measurement of $\sigma_{\rm v}$ propagates to a 10\% uncertainty in $\lambda$.
Beyond the measurement uncertainty in $\sigma_{\rm v}$, projection uncertainties and degeneracies are present in the interpretation of the measurement, the model uncertainty. In particular, the mass-anisotropy degeneracy limits the precision, so only spatially resolved kinematics observations are able to break this secondary, but relevant, degeneracy \citep[e.g.,][]{Yildirim:2020, BirrerTreu:2021}. Constraints on the radial extent of the mass profile with spatially resolved kinematics are possible. Equation~\ref{eqn:v_disp_mst} is, however, applicable for the covariant uncertainties among multiple measurements or integral-field unit spectroscopy. A forecast utilizing kinematic measurements of ground- and space-based facilities on a larger sample of lenses within the same assumptions as \cite{Birrer:2020} is presented by \cite{BirrerTreu:2021}.

\subsubsection{Absolute lensing magnifications}
The alternative to kinematics, and key element in the exploration in this work, are absolute magnification constraints \citep[][]{Kolatt:1998, Foxley-Marrable:2018}. Absolute lensing magnifications, $\mu$, change under an MST by
\begin{equation} \label{eqn:mag_mst}
    \mu_{\lambda} = \lambda^{-2} \mu.
\end{equation}
A fractional uncertainty in the lensing magnification propagates to a fractional uncertainty in the MST as
\begin{equation} \label{eqn:mu_mst_propagation}
    \frac{\delta \lambda}{\lambda} = -0.5\frac{\delta \mu}{\mu} = -0.5\left[ \frac{\delta \mu_{\rm obs}}{\mu_{\rm obs}} - \frac{\delta \mu_{\rm model}}{\mu_{\rm model}} \right].
\end{equation}

The observed magnification $\mu_{\lambda}$ is the ratio
\begin{equation} \label{eqn:magnification_measurement}
    \mu_{\rm obs} = \frac{F_{\rm obs}}{F_{\rm unl}},
\end{equation}
where $F_{\rm obs}$ is the observed flux of an image, and $F_{\rm unl}$ is the unlensed apparent brightness of the object in the absence of the lensing effect. While measuring the observed flux of a lensed object is achieved to sub-percent precision on a regular basis, a lensing-independent measurement of $\mu_\lambda$ requires, in addition, knowledge of the unlensed apparent brightness of the object in the same observational band as the measurement.
We stress that the measurement of the lensing magnification does not require knowledge or calibration of the absolute luminosity, which is a key requirement in measuring $H_0$ with SNe~Ia \citep[e.g.,][]{Riess:2019, Freedman:2019}. Only the probability distribution function of the apparent magnitude of the source at the redshift of the source is required.

The estimation of the MST scaling for a given lens model relevant for the time-delay prediction, and thus the measurement of $H_0$, requires, in addition, an accurate lensing magnification prediction in accordance to the Fermat potential prediction. While the time-delay prediction is less susceptible to small scale model inaccuracies\footnote{See \cite{Gilman:2020time_delays} for the impact of unresolved small scale dark matter structure on the predicted time delays, resulting in a scatter of about 2.5\% on an individual lens.} as it relies only on an accurate lensing potential, the local magnification is impacted more significantly, as it relies on the second-order differentials of the potential. In addition to small scale dark matter structure, both along the line of sight (LOS) and within the main deflector as substructure, stellar microlensing is an additional source of lensing magnifications for sources of the size of exploding SNe \citep[e.g.,][]{Dobler:2006, Foxley-Marrable:2018, Suyu:2020}.

We can approximately separate the different components entering the local magnification prediction into a smooth macro-model component $\mu_{\rm macro}$ and an additional perturbation by dark matter structure on milli-arcsecond scales, $\Delta\mu_{\rm milli}$, and stellar microlensing, $\Delta\mu_{\rm micro}$, as\footnote{Magnification effects are, in general, not additive. We justify the approximation by the different scales of macro-, milli-, and micro-lensing.}
\begin{equation}\label{eqn:magnification_split}
    \mu_{\rm local} \approx \mu_{\rm macro} + \Delta\mu_{\rm milli} + \Delta\mu_{\rm micro}.
\end{equation}

Milli-lensing depends on the halo substructure and on the line-of-sight abundances of small field halos. Stellar micro-lensing depends on the local projected stellar surface density and can vary significantly from lens to lens and from image position to image position.

The relative difference in $\lambda$ for either an infinitesimal change in the apparent unlensed magnitude, $\delta F_{\rm unl}$, a change in the lensed observed flux, $\delta F_{\rm obs}$, the model predicted magnification, $\delta\mu_{\rm model}$, or the physical cause of local milli (micro) lensing, $\delta \mu_{\rm milli}$ ($\delta\mu_{\rm micro}$), while keeping all other quantities fixed, can be expressed as (Eqn.~\ref{eqn:mu_mst_propagation},~\ref{eqn:magnification_measurement},~\ref{eqn:magnification_split})

\begin{multline}\label{eqn:error_differential}
    \frac{\delta \lambda}{\lambda}
    = 0.5 \left[ \frac{\delta F_{\rm unl}}{F_{\rm unl}} - \frac{\delta F_{\rm obs}}{F_{\rm obs}} \right. \\ + \left. \frac{\delta\mu_{\rm macro}}{\mu} 
    +  \frac{\delta\Delta\mu_{\rm milli}}{\mu} + \frac{\delta\Delta\mu_{\rm micro}}{\mu} \right].
\end{multline}

In words, while keeping all other parameters fixed, an increase in $F_{\rm unl}$ leads to an increase in $\lambda$, an increase in $F_{\rm obs}$ leads to a decrease in $\lambda$, an increase in the lensing magnifications $\mu_{\rm macro}$, $\Delta\mu_{\rm milli}$ and $\Delta\mu_{\rm milli}$ leads to an increase in $\lambda$. On the other hand, errors in the measurement or estimation of these quantities result in shifts of $\lambda$ in the opposite direction.

The intrinsically small scatter of Type~Ia supernovae is a well suited population to constrain the MST. An intrinsic scatter of 10\% in the peak brightness after lightcurve width and color corrections \citep{Scolnic:2018} allows one, at least in principle, to constrain the MST to 5\% in the absence of other uncertainties. Thus, glSNe are not only able to provide precise time-delay measurements due to their transient and well characterized nature, but at the same time Type~Ia or any other standardizable form of SNe allows one to constrain the currently dominating error budget of time-delay cosmography, the MST.

The constraints on the MST rely on precise and accurate determinations of all the parameters listed in Equation~\ref{eqn:error_differential}.
The uncertainty in apparent unlensed brightness of SNe $F_{\rm unl}$, milli-lensing $\Delta\mu_{\rm milli}$ contribution and the microlensing effect $\Delta\mu_{\rm micro}$ are the dominant uncertainty components in constraining the MST.
Systematic limitations in the usage of glSNe relate to dust extinction impacting the flux measurement $F_{\rm obs}$, and selection effects related to milli- and micro-lensing.
We will review limitations and systematics of glSNE in breaking the MST in Section~\ref{sec:discussion}.

\section{Methodology} \label{sec:method}
In this section, we describe the methodology to measure $H_0$ from a set of glSNe by constraining the MST with the apparent magnitude distribution of an unlensed SNe sample.
We layout the model assumptions and define the hyper-parameters governing the cosmological expansion, SNe brightness distribution, and the mass profiles of the lensing galaxies. Furthermore, we detail the implementation of the likelihood for the different observations that allow us to efficiently perform a joint hierarchical sampling of the posteriors. We describe the SNe~Ia population assumptions and analysis in Section~\ref{sec:sne_analysis}, and the glSNe population assumptions and analysis in Section~\ref{sec:deflector_population}. Separately, we discuss the impact and the treatment of LOS structure in Section~\ref{sec:los}. In Section~\ref{sec:hierarchical_analysis}, we state the joint hierarchical inference problem based on the previous parts of this section. In Section~\ref{sec:analytic_error_propagation}, we provide an approximate analytical error propagation.
The methodology presented here, in terms of parameterization and likelihood calculation, is implemented in the open-source software \textsc{hierArc}.

\subsection{Supernovae of Type~Ia population}\label{sec:sne_analysis}
We focus in this work on the SNe~Ia population. Here we describe the SNe~Ia magnitude--redshift relation and the likelihood for the unlensed SN~Ia sample in Section~\ref{sec:app_vs_bol_brightness}. In Section~\ref{sec:sne_model_param}, we then state the specific model and likelihood assumptions for the combined lensed and unlensed SNe~Ia samples that we implement in this work.

\subsubsection{Unlensed SN~Ia sample}\label{sec:app_vs_bol_brightness}

SNe~Ia can be standardized as precise (relative) distance indicators. This involves well-known corrections for their luminosity-width and luminosity-color relations \citep{Guy:2007js, Guy:2010}. In addition, the SN~Ia inferred luminosity needs to be corrected for its dependence on the host galaxy properties (e.g. stellar mass) \citep[refer to ][for details of the standardization procedure and bias corrections]{Scolnic:2018}.
The distance modulus, $\mu_{\rm B}$, relates the standardized apparent magnitude of an unlensed SN~Ia, denoted as $m_{\rm B}^*$, with the absolute magnitude $M_{\rm B}$ as
\begin{equation}
    m_{\rm B}^{*} - M_{\rm B} = \mu_{\rm dist} = 5 \log \left(D_L(z)\right) + 25,
    \label{eq:mu_mb}
\end{equation}
where $D_L(z)$ is the luminosity distance from the observer to the redshift of the SNe, and $\mu_{\rm dist}$ is the distance modulus.

There are several large samples of the cosmological SNe~Ia in the literature, e.g. Pantheon \citep{Scolnic:2018}, Joint Light-curve Analysis \citep[JLA; ][]{Betoule:2014}, or the Dark Energy Survey \citep[DES; ][]{DES_SNe:2019}. 
For our analyses we use the largest, up-to-date compilation, i.e. the Pantheon SN~Ia dataset \citep{Scolnic:2018} 
For these SN~Ia samples,  covariances in the calibration parameters $\boldsymbol{\xi}_{\rm sys}$ and their evolutionary trends need to be taken into account.

Lensing magnifications, be it weak lensing from large-scale structures, strong lensing from massive deflectors, or micro-lensing from stars, change the flux or apparent magnitude by

\begin{align}\label{eqn:magnification_flux_magnitude}
    F_{\mu} = \mu F && m_{\mu} = m - 2.5 \log_{10}(\mu),
\end{align}
where $\mu$ is the unsigned absolute magnification.
For a single SN, we can formally write down the likelihood of an observed peak brightness $F_{\rm obs}$ given a luminosity distance $D_{L}$, lensing magnification $\mu$, and absolute magnitude $M_{\rm B}$ as the likelihood of the data given a flux prediction $F_{\rm model}$ while marginalizing over calibration and other uncertainties, such as dust extinction, or uncertainty in the peak time, as

\begin{multline}
    \mathcal{L}(F_{\rm obs} \mid D_{L}, \mu, M_{\rm b}) = \\
    \int \mathcal{L}(F_{\rm obs} \mid F_{\rm model}) p(F_{\rm model} \mid D_{L}, \mu, M_{\rm b}, \boldsymbol{\xi}_{\rm sys})p(\boldsymbol{\xi}_{\rm sys}) d\boldsymbol{\xi}_{\rm sys}.
\end{multline}

For a sample of SNe, denoted as $\mathcal{D}_{\rm SNe}$, different procedures and models have been employed for the parameterization, calibration, and marginalization of systematic errors on the population level and we refer to the relevant work for details \citep[e.g.,][]{Betoule:2014, Scolnic:2018, DES_SNe:2019}. For this work, when combining such a supernovae sample with glSNe, we require the likelihood $\mathcal{L}(\mathcal{D}_{\rm SNe} \mid \boldsymbol{\pi}, \boldsymbol{\xi}_{\rm SNe})$ of the global data set given the cosmological prediction of the luminosity distances with parameters $\boldsymbol{\pi}$ and the intrinsic brightness distribution of the SNe population $\boldsymbol{\xi}_{\rm SNe}$.

To facilitate the evaluation of the likelihood, for example, \cite{Scolnic:2018} compressed the marginalization in a Gaussian covariance matrix across all the measured apparent magnitudes in the SNe sample as 

\begin{multline} \label{eqn:sne_sample}
    \mathcal{L}(\mathcal{D}_{\rm SNe} \mid \boldsymbol{\pi}, \boldsymbol{\xi}_{\rm SNe})\\ = \frac{1}{\sqrt{(2\pi)^{n_{\rm sn}} {\rm det}(\mathbf{\Sigma_{\rm cov}})}} 
    \exp \left[-\frac{1}{2}\boldsymbol{\Delta m}^{\rm T} \mathbf{\Sigma_{\rm cov}^{-1}} \boldsymbol{\Delta m}\right],
\end{multline}
where $\boldsymbol{\Delta m}$ is the difference in the observed and predicted apparent magnitude of a non-evolving intrinsic mean brightness of the SNe population, $n_{\rm sn}$ is the length of the data vector, and $\mathbf{\Sigma_{\rm cov}}$ is the error covariance matrix when marginalized over the systematics variables in Gaussian form.

\subsubsection{Model parameterization and assumptions with pivot magnitude}\label{sec:sne_model_param}

We assume, for simplicity of this work, that the intrinsic peak brightness distribution is redshift independent. This is typically assumed in cosmological analyses with SNe~Ia, based on comparisons of spectroscopic and photometric properties of local and high-$z$ SNe~Ia \citep[e.g.][and other studies of high signal-to-noise data of high-$z$ SNe~Ia]{petrushevska:2017}. 

To obtain the absolute luminosity of an unlensed SN~Ia at the redshift of the glSNe in our sample would require an independent calibration of the absolute luminosity distance, e.g. as done for the distance ladder. However, since we want to derive a relative magnification at the lensed source redshift, we can use the apparent magnitude of the unlensed SNe~Ia from the cosmological sample and infer it at the redshift of the lensed SN. For this, we replace using an $M_{\rm B}$ term with a apparent magnitude at a specific redshift, $z_{\rm pivot}$, $m_{\rm p}$
\begin{equation}\label{eqn:pivot_magnitude}
    m_{\rm sn}(m_{\rm p}, z) = m_{\rm p} + 5 \left[\log_{10}D_L(z) - \log_{10}D_L(z_{\rm pivot})\right].
\end{equation}
This parameterization results in a likelihood which is only dependent on relative distance ratios without the need of external data or constraints on top of a population of observed peak brightness of SNe.
We describe the intrinsic distribution of apparent peak brightness at the pivot redshift $p(m_{\rm p})$ by a Gaussian in astronomical magnitude space with a mean $\overline{m}_{\rm p}$ and width $\sigma(m_{\rm p})$.

With these simplifications, we can write the likelihood for a single SNe as

\begin{multline}
    \mathcal{L}(F_{\rm obs} \mid D_{L}, \mu, \overline{m}_{\rm p}, \sigma(m_{\rm p}))\\
    = \int \frac{1}{\sqrt{2\pi} \sigma_{\rm obs}} \exp\left[-\frac{\left(F_{\rm obs} - \mu F'(m_{\rm p}')\right)^2}{2\sigma^2_{\rm obs}} \right] \\
    \times p(m_{\rm p}'\mid \overline{m}_{\rm p}, \sigma(m_{\rm p}))dm_{\rm p}',
\end{multline}
where $F'(m_{\rm p}')$ is the shorthand form of the model predicted flux given an apparent magnitude $m_{\rm sn}$ calculated by Equation~\ref{eqn:pivot_magnitude} from $m_{\rm p}'$ and the luminosity distance ratio, and then turned into flux units of the observations while considering the lensing magnification (Eqn.~\ref{eqn:magnification_flux_magnitude}). $\sigma_{\rm obs}$ is the Gaussian error in the flux measurements and the term $p\left(m_{\rm p}' \mid \overline{m}_{\rm p}, \sigma(m_{\rm p})\right)$ in the equation above describes the likelihood of a specific pivot magnitude to be drawn from the Gaussian distribution $\mathcal{N}\left(\overline{m}_{\rm p}, \sigma(m_{\rm p})\right)$.

For an ensemble of SNe characterized with the likelihood of Equation~\ref{eqn:sne_sample}, the marginalization over a Gaussian distribution in $m_{\rm p}$ is analytic and can directly be folded in the error covariance matrix as
\begin{equation}
    \mathbf{\Sigma'_{\rm cov}} = \mathbf{\Sigma_{\rm cov}} + {\rm diag}\left(\sigma^2(m_{\rm p})\right)
\end{equation}
and the marginalized likelihood is given by

\begin{multline} \label{eqn:sne_sample_marg}
    \mathcal{L}(\mathcal{D}_{\rm SNe} \mid \boldsymbol{\pi}, \overline{m}_{\rm p}, \sigma(m_{\rm p})) 
    \\= \frac{1}{\sqrt{(2\pi)^{n_{\rm sn}} {\rm det}(\mathbf{\Sigma'_{\rm cov}})}}\exp \left[-\frac{1}{2}\boldsymbol{\Delta m}^{\rm T} \mathbf{\Sigma_{\rm cov}^{'-1}} \boldsymbol{\Delta m}\right].
\end{multline}

\subsection{Deflector population}\label{sec:deflector_population}
We first discuss general considerations about the deflector parameterization and the necessary degrees of freedom to allow for an accurate recovery of the time-delay prediction (Section~\ref{sec:glSNe_deflector}). We then formulate the inference problem and the general form of the joint likelihood of the imaging data, time delays, and lensed SNe peak brightness observations provided by a glSNe (Section~\ref{sec:glSNe_inference}). Lastly, we provide a Gaussian approximation of the likelihood for a fast marginalization and efficient evaluation (Section \ref{sec:glSNe_gaussian_approx}).

\subsubsection{Deflector parameterization and measurements}\label{sec:glSNe_deflector}
The uncertainty in the deflector mass distribution dominates the current error budget in the $H_0$ inference, a statement directly reflecting the MST.
A popular model describing strong gravitational lensing imaging data on galaxy-scale lenses is the power-law elliptical mass distribution \citep[PEMD;][]{Barkana:1998, Tessore:2015} combined with an external shear component. The popularity of the PEMD+shear model is a consequence of its ability to describe the data sufficiently well while keeping the degrees of freedom in the deflector model to a computationally affordable number.

Although considered simplistic, the PEMD+shear model's degrees of freedom can describe the primary azimuthal and radial observables. However, the observable in the radial direction are related to the third order differential of the lensing potential, while the parameterization of the PEMD profile explicitly assumes a one-to-one connection between the observable invariant quantity and the mass density at the position of the Einstein ring, leading to over-constrained mass profiles and potentially biased inferences in the radial density profile, and subsequently $H_0$ \citep[see e.g.][]{Kochanek:2002, Sonnenfeld:2018, Kochanek:2020, Kochanek:2021, Birrer:2020, Birrer:2021arcs}.

To mitigate possible over-constraints on the internal mass density profile, \cite{Birrer:2020} added an additional degree of freedom with an MST on top of the PEMD+shear profiles of the TDCOSMO sample. A PEMD+shear+MST profile has the adequate degrees of freedom at and around the Einstein ring, where the multiple images appear, to estimate the relative Fermat potential. Higher-order differentials are subdominant in the effect on the predicted time delays \citep[e.g.,][]{Sonnenfeld:2018}.
In addition, since the constraints on the MST from the lensing magnification are directly derived at the region relevant for the time-delay prediction, potential inadequacies of the PEMD+shear+MST profile further outwards or towards the center of the deflector do not impact the accuracy in the inferred time-delay prediction, and thus $H_0$ measurement.

In this work, we assume that the population of lenses can be described by a PEMD+shear+MST profile. Furthermore, we assume that the PEMD+shear parameters can be measured accurately for each lens individually from the imaging data without population covariances, and that the MST parameter $\lambda$ transforming the internal density profile of the main deflector, denoted as $\lambda_{\rm int}$, follows a Gaussian distribution with mean $\lambda_{\rm int}$ and sigma $\sigma(\lambda_{\rm int})$. We refer to Section~\ref{sec:los} for a discussion on different MST components. We highlight that there can be physical covariances between the PEMD parameters and $\lambda_{\rm int}$, as well as among the physical projected scale and $\lambda_{\rm int}$. A possible physical projection dependence has been accounted for by \cite{Birrer:2020} with an explicit parameterization of $\lambda_{\rm int}$ as a function of the ratio of deflector half-light radius relative to the Einstein radius. In this work, for the purpose of providing a forecast, we do not include secondary dependencies and covariances of $\lambda_{\rm int}$, and instead refer to \cite{Birrer:2020} for the radial dependence as well as to \cite{Wagner-Carena:2021} for a general treatment of lens model hyper-parameter inferences within a hierarchical framework.

\subsubsection{glSNe inference}\label{sec:glSNe_inference}

From the imaging data, $\boldsymbol{I}$, we can measure the lens model parameters within our model assumptions, $\boldsymbol{\xi}_{\rm pl}$, which in turn provide the Fermat potential differences between the multiple images $\boldsymbol{\Delta \tau}_{\rm pl}$ and the lensing magnifications $\boldsymbol{\mu}_{\rm pl}$ at the position of the appearances of the glSN. With measured relative time delays $\boldsymbol{\Delta t}$ and a model providing values for $\boldsymbol{\Delta \tau}_{\rm pl}$, $\lambda$, and $D_{\Delta t}$, we can predict the time delay and evaluate the time-delay likelihood of the data given the model. From the same lens model, we can compute the likelihood of the measured glSNe brightness $\boldsymbol{F}$ given the model prediction of $\boldsymbol{\mu}_{\rm pl}$, $\lambda$, and $m_{\rm sn}$.

The joint likelihood $\mathcal{L}(\boldsymbol{I}, \boldsymbol{\Delta t}, \boldsymbol{F} \mid D_{\Delta t}, m_{\rm sn}, \lambda)$ of the imaging, time delay, and flux measurements -- given the relevant parameters of the hierarchical inference, $D_{\Delta t}$, $m_{\rm sn}$, and $\lambda$ -- can be written as product of the likelihoods of the different independent data sets

\begin{multline} \label{eqn:lens_likelihood}
    \mathcal{L}(\boldsymbol{I}, \boldsymbol{\Delta t}, \boldsymbol{F} \mid D_{\Delta t}, m_{\rm sn}, \lambda) = \int \mathcal{L}(\boldsymbol{I} \mid \boldsymbol{\Delta\tau}_{\rm pl}, \boldsymbol{\mu}_{\rm pl}) \\\times
     \mathcal{L}(\boldsymbol{\Delta t} \mid D_{\Delta t}, \lambda, \boldsymbol{\Delta\tau}_{\rm pl}) \mathcal{L}(\boldsymbol{F} \mid m_{\rm sn}, \lambda, \boldsymbol{\mu}_{\rm pl})\\
     \times p(\boldsymbol{\Delta\tau}_{\rm pl}, \boldsymbol{\mu}_{\rm pl})
    d\boldsymbol{\Delta\tau}_{\rm pl} d\boldsymbol{\mu}_{\rm pl},
\end{multline}
where we explicitly marginalized over the magnification and Fermat potential parameters $\boldsymbol{\mu}_{\rm pl}$ and $\boldsymbol{\tau}_{\rm pl}$. To describe the imaging data $\boldsymbol{I}$ and to compute the likelihood at the pixel level, we require a lens model $\boldsymbol{\xi}_{\rm pl}$ and a model of all the light components $\boldsymbol{\xi}_{\rm light}$. We can describe the imaging likelihood and prior product on $\boldsymbol{\Delta\tau}_{\rm pl}$ and $\boldsymbol{\mu}_{\rm pl}$ as

\begin{multline}\label{eqn:imaging_likelihood}
    \mathcal{L}(\boldsymbol{I} \mid \boldsymbol{\Delta\tau}_{\rm pl}, \boldsymbol{\mu}_{\rm pl}) p(\boldsymbol{\Delta\tau}_{\rm pl}, \boldsymbol{\mu}_{\rm pl})
    \\
    = \int \mathcal{L}(\boldsymbol{I} \mid \boldsymbol{\xi}_{\rm pl}, \boldsymbol{\xi}_{\rm light}) p(\boldsymbol{\Delta\tau}_{\rm pl}, \boldsymbol{\mu}_{\rm pl}\mid \boldsymbol{\xi}_{\rm pl}) p(\boldsymbol{\xi}_{\rm pl}, \boldsymbol{\xi}_{\rm light})
    d\boldsymbol{\xi}_{\rm pl}\boldsymbol{\xi}_{\rm light}
    \\ = 
    \int \mathcal{L}(\boldsymbol{I} \mid \boldsymbol{\xi}_{\rm pl}, \boldsymbol{\xi}_{\rm light}) p(\boldsymbol{\xi}_{\rm pl}, \boldsymbol{\xi}_{\rm light}) \left|\frac{\partial(\boldsymbol{\Delta\tau}_{\rm pl}, \boldsymbol{\mu}_{\rm pl})}{\partial ( \boldsymbol{\xi}_{\rm pl}, \boldsymbol{\xi}_{\rm light})}\right|^{-1} d\boldsymbol{\xi}_{\rm pl}\boldsymbol{\xi}_{\rm light}.
\end{multline}
Here, $\Delta\tau(\boldsymbol{\xi}_{\rm pl})$ and $\boldsymbol{\mu}(\boldsymbol{\xi_{\rm pl}})$ are unique functions of $\boldsymbol{\xi}_{\rm pl}$, and the $\left|{\partial(\boldsymbol{\Delta\tau}_{\rm pl}, \boldsymbol{\mu}_{\rm pl})}/{\partial ( \boldsymbol{\xi}_{\rm pl}, \boldsymbol{\xi}_{\rm light})}\right|$ is the Jacobian determinant.
This means that the likelihood and prior product of Equation~\ref{eqn:imaging_likelihood} can be computed by sampling $\boldsymbol{\xi}_{\rm pl}$ from the posterior $\mathcal{L}(\boldsymbol{I} \mid \boldsymbol{\xi}_{\rm pl}, \boldsymbol{\xi}_{\rm light})p(\boldsymbol{\xi}_{\rm pl}, \boldsymbol{\xi}_{\rm light})$, and evaluating for the posterior sample of the quantities $\boldsymbol{\Delta\tau}_{\rm pl}(\boldsymbol{\xi}_{\rm pl})$ and $\boldsymbol{\mu}_{\rm pl}(\boldsymbol{\xi}_{\rm pl})$.
For the modeling choices and posterior sampling when marginalizing over complex source structure, we refer to previous work \citep[e.g.,][]{Suyu:2009, Birrer:2015}.

\subsubsection{Gaussian likelihood approximation}\label{sec:glSNe_gaussian_approx}
Until this point in this subsection, we did not make any assumption on the form of the likelihood (Eqn.~\ref{eqn:lens_likelihood}) nor on the shape of the imaging modeling posteriors (Eqn.~\ref{eqn:imaging_likelihood}). To facilitate the calculation of the likelihood in Equation~\ref{eqn:lens_likelihood}, we approximate the likelihood in Gaussian form. In particular, we write the imaging likelihood from Equation~\ref{eqn:imaging_likelihood} as

\begin{multline} \label{eqn:imaging_likelihood_gaussian}
    \mathcal{L}(\boldsymbol{I} \mid \boldsymbol{\Delta\tau}_{\rm pl}, \boldsymbol{\mu}_{\rm pl}) p(\boldsymbol{\Delta\tau}_{\rm pl}, \boldsymbol{\mu}_{\rm pl})\\
      \approx \frac{1}{\sqrt{(2\pi)^{n_{\Delta\tau\mu}}\det( \mathbf{\Sigma}_{\Delta\tau\mu})}} \exp \left[-\frac{1}{2}\boldsymbol{\Delta}^{\rm T}_{\Delta\tau\mu}\mathbf{\Sigma}^{-1}_{\Delta\tau\mu}\boldsymbol{\Delta}_{\Delta\tau\mu} \right],
\end{multline}
where $\boldsymbol{\Delta}_{\Delta\tau\mu} \equiv (\boldsymbol{\Delta\tau} - \boldsymbol{\Delta\tau}_0, \boldsymbol{\mu} - \boldsymbol{\mu}_0)$ with $(\boldsymbol{\Delta\tau}_0, \boldsymbol{\mu}_0)$ being the maximum likelihood estimator, $n_{\Delta\tau\mu}$ is the length of the vector $\boldsymbol{\Delta}_{\Delta\tau\mu}$, and $\mathbf{\Sigma}_{\Delta\tau\mu}$ is the error covariance matrix describing the Gaussian uncertainties in the measurement from the imaging data.

The Gaussian form of the time-delay likelihood $\mathcal{L}(\boldsymbol{\Delta t} \mid D_{\Delta t}, \lambda, \boldsymbol{\Delta\tau}_{\rm pl})$ of Equation~\ref{eqn:lens_likelihood} reads

\begin{multline} \label{eqn:td_likelihood_gaussian}
    \mathcal{L}(\boldsymbol{\Delta t} \mid D_{\Delta t}, \lambda, \boldsymbol{\Delta\tau}_{\rm pl})\\
    \approx \frac{1}{\sqrt{(2\pi)^{n_{\Delta t}}\det( \mathbf{\Sigma}_{\Delta t})}} \exp \left[-\frac{1}{2}\boldsymbol{\Delta}^{\rm T}_{\Delta t}\mathbf{\Sigma}^{-1}_{\Delta t}\boldsymbol{\Delta}_{\Delta t} \right],
\end{multline}
where $n_{\Delta t}$ are the number of relative time delay measurements, $\mathbf{\Sigma}_{\Delta t}$ is the relative time-delay measurement error covariance matrix, and $\boldsymbol{\Delta}_{\Delta t}$ is the difference between the predicted time delay (Eqn.~\ref{eqn:time_delay} including an MST term of Eqn.~\ref{eqn:time_delay_mst}) and measured time delay $\boldsymbol{\Delta t}$.

The Gaussian form of the flux amplitude likelihood $\mathcal{L}(\boldsymbol{F} \mid m_{\rm sn}, \lambda, \boldsymbol{\mu}_{\rm pl})$ of Equation~\ref{eqn:lens_likelihood} is
\begin{multline} \label{eqn:flux_likelihood_gaussian}
    \mathcal{L}(\boldsymbol{F} \mid m_{\rm sn}, \lambda, \boldsymbol{\mu}_{\rm pl}) \\
    \approx \frac{1}{\sqrt{(2\pi)^{n_F}\det( \mathbf{\Sigma}_{F})}} \exp \left[-\frac{1}{2}\boldsymbol{\Delta}^{\rm T}_{F}\mathbf{\Sigma}^{-1}_{F}\boldsymbol{\Delta}_{F} \right],
\end{multline}
where $n_{F}$ is the number of flux measurements, $\mathbf{\Sigma}_{F}$ is the glSNe flux measurement error covariance matrix, and $\boldsymbol{\Delta}_{F}$ is the difference between the predicted peak flux of the SNe and measured flux $\boldsymbol{F}$.

The marginalized likelihood of Equation~\ref{eqn:lens_likelihood} with the 
Gaussian approximations for the individual likelihood components (Eqn.~\ref{eqn:imaging_likelihood_gaussian}, \ref{eqn:td_likelihood_gaussian}, \ref{eqn:flux_likelihood_gaussian}) is a Gaussian integral.

We can join the data vector of the time delays $\boldsymbol{\Delta t}$ and fluxes $\boldsymbol{F}$ as $\boldsymbol{d}_{\Delta t F} \equiv (\boldsymbol{\Delta t}, \boldsymbol{F})$ and write the joint measurement covariance matrix as
\begin{equation}
    \mathbf{\Sigma}_{\rm data} = \begin{bmatrix}
\mathbf{\Sigma}_{\Delta t} & \mathbf{0}\\
\mathbf{0} & \mathbf{\Sigma}_{F}
\end{bmatrix}.
\end{equation}
In this forecast, we assume no covariant measurement uncertainties between the time delays and the micro-lensing impact on the magnification. It has been shown that the micro-lensing effect on time-delay measurements can be mitigated with early phase multi-color light curves when the micro-lensing effect is achromatic \citep{Goldstein:2018}.

At the same time, we can transform the covariance matrix of the imaging posteriors $\mathbf{\Sigma}_{\Delta \tau\mu}$ into the data vector space, resulting in

\begin{multline}
    \mathbf{\Sigma}_{\rm model} = (\lambda D_{\Delta t}c^{-1} \boldsymbol{1}_{n_{\Delta \tau}}, \lambda^{-2} m_{\rm sn} \boldsymbol{1}_{n_{\mu}} )\mathbf{\Sigma}_{\Delta \tau\mu} \\ (\lambda D_{\Delta t} c^{-1} \boldsymbol{1}_{n_{\Delta \tau}}, \lambda^{-2} m_{\rm sn} \boldsymbol{1}_{n_{\mu}})^{\rm T},
\end{multline}
where $\boldsymbol{1}_{n_{\Delta \tau}}$ and $\boldsymbol{1}_{n_{\mu}}$ are vectors of size $n_{\Delta \tau}$ and ${n_{\mu}}$, respectively, with 1 at each element.
The joint likelihood of Equation~\ref{eqn:lens_likelihood} is then given by

\begin{multline}\label{eqn:gauss_lens_likelihood}
    \mathcal{L}(\boldsymbol{I}, \boldsymbol{\Delta t}, \boldsymbol{F}\mid D_{\Delta t}, m_{\rm sn}, \lambda)\\ = 
    \frac{1}{\sqrt{(2\pi)^{n_{\Delta t F}}\det( \mathbf{\Sigma}_{\rm tot})}} \exp \left[-\frac{1}{2}\boldsymbol{\Delta}^{\rm T}_{\Delta t F}\mathbf{\Sigma}^{-1}_{\rm tot}\boldsymbol{\Delta}_{\Delta t F} \right],
\end{multline}
with $n_{\Delta t F} \equiv n_{\Delta t} + n_{F}$, and the total error covariance matrix being the sum of the measurement covariance and the marginalized uncertainty in the model
\begin{equation}
    \mathbf{\Sigma}_{\rm tot} = \mathbf{\Sigma}_{\rm data} + \mathbf{\Sigma}_{\rm model},
\end{equation}
and $\boldsymbol{\Delta}_{\Delta t F}$ being the difference of the data vector $\boldsymbol{d}_{\Delta t F}$ and the model prediction.

\subsection{LOS mass distribution}\label{sec:los}

Mass over- or under-densities along the LOS of the strong lensing system cause, to first order, shear and convergence perturbations. Reduced shear distortions do have a measurable imprint on the azimuthal structure of the strong lensing system \citep[see e.g.,][]{Birrer:2021arcs} while the convergence component of the LOS, denoted as $\kappa_{\rm ext}$ is equivalent to an MST, and thus not directly measurable from imaging data. The total MST, the relevant transform to constrain for an accurate cosmography and $H_0$ measurement, is the product of the internal and external MST \citep[e.g.,][]{SchneiderSluse:2013, Birrer:2016, Birrer:2020}
\begin{equation}\label{eqn:lambda_combined}
    \lambda = (1-\kappa_{\rm ext}) \times \lambda_{\rm int}.
\end{equation}

The lensing kernel impacting the linear distortions, both shear and $\kappa_{\rm ext}$ is different from the standard weak lensing kernel \citep{McCully:2014, McCully:2017, Birrer:2017los, Birrer:2020, Fleury:2021}. The lensing kernel can be described as the product of three different angular diameter distances entering $D_{\Delta t}$ in Equation~\ref{eqn:ddt_definition} \citep{Birrer:2020, Fleury:2020}, and thus $\kappa_{\rm ext}$ can be described as the product of the individual kernels entering Equation \ref{eqn:ddt_definition} as
\begin{equation}\label{eqn:mst_combined}
    1 - \kappa_{\rm ext} = \frac{(1 - \kappa_{\rm d})(1 - \kappa_{\rm s})}{1 - \kappa_{\rm ds}},
\end{equation}
where $\kappa_{\rm d}$ is the weak lensing effect from the observer to the deflector, $\kappa_{\rm s}$ from the observer to the source, and $\kappa_{\rm ds}$ from the deflector to the source, respectively \citep{Birrer:2020}.
Alternatively, but equivalently, the kernel can be described in the multi-plane formalism with the main deflector included, while keeping the Born approximation in between \citep[e.g.,][]{Birrer:2017los, Fleury:2021}.

The LOS lensing contribution can be estimated by tracers of the large-scale structure, either using galaxy number counts \citep[e.g.,][]{Greene:2013, Rusu:2017}, or weak-lensing measurements \citep{Tihhonova:2018}. These measurements, paired with a cosmological model including a galaxy--halo connection are able to constrain the probability distribution of $\kappa_{\rm ext}$ to few per cent per LOS.

For an accurate measurement of $H_0$, the combined internal and external MST of Equation~\ref{eqn:mst_combined} is required. Since glSNe magnification is directly probing the combined $\lambda$, the LOS contribution effectively only adds a scatter in the inference and an accurate overall population selection function is not required \citep[see][for the same argument using kinematics to break the MST]{Birrer:2020}.
The overall lensing selection function is only relevant when demanding a physical interpretation of the internal and external contributions separately.

In this work, for practical simplicity but without impact on expected biases or uncertainty budget, we assume a Gaussian scatter in $\kappa_{\rm ext}$ of 0.03 with a population mean at zero along the LOS's of the lenses.

\subsection{Hierarchical analysis and sampling}\label{sec:hierarchical_analysis}

Our goal is to jointly infer and marginalize over population hyper-parameters in the SNe distribution, lensing deflector profiles, and cosmological parameters, given the joint data set of lensed and unlensed SNe, and MST-invariant lensing quantities from imaging data.
We follow the same approach as \cite{Birrer:2020}, except that we add the SNe magnification likelihood instead of the stellar kinematic one, and as an external data set we are using a sample of unlensed SNe instead of a sample of galaxy--galaxy lenses with measured kinematics.

We want to calculate the probability of the cosmological parameters, $\boldsymbol{\pi}$, given the joint data set, $p(\boldsymbol{\pi} \mid \{\mathcal{D}^{i}_{\rm L} \}_{N}, \mathcal{D}_{\rm SNe})$, where $\mathcal{D}^{i}_{\rm L}$ is the data set of an individual strong lens (including imaging data, time-delay measurements, SNe flux measurement, and LOS properties), $N$ is the total number of lenses in the sample, and $\mathcal{D}_{\rm SNe}$ is a SNe data set.

In addition to $\boldsymbol{\pi}$, we introduce $\boldsymbol{\xi}_{\rm pop}$ that incorporates all the additional population-level model parameters not yet marginalized over the individual data sets including their covariant impact on the likelihoods of individual lenses. Using Bayes' rule and considering that the data of each individual lens $\mathcal{D}_{i}$ is independent, we can write:

\begin{multline} \label{eqn:full_inference}
    p(\boldsymbol{\pi} \mid \{\mathcal{D}^{i}_{\rm L} \}_{N}, \mathcal{D}_{\rm SNe}) \propto \mathcal{L}(\{\mathcal{D}^{i}_{\rm L} \}_{N}, \mathcal{D}_{\rm SNe}\mid \boldsymbol{\pi}) p(\boldsymbol{\pi}) \\
    = \int \mathcal{L}(\{\mathcal{D}^{i}_{\rm L} \}_{N}, \mathcal{D}_{\rm SNe}\mid \boldsymbol{\pi}, \boldsymbol{\xi}_{\rm pop})p(\boldsymbol{\pi}, \boldsymbol{\xi}_{\rm pop}) d \boldsymbol{\xi}_{\rm pop} \\
    = \int \left[ \prod_i^N \mathcal{L}(\mathcal{D}^{i}_{\rm L}\mid \boldsymbol{\pi}, \boldsymbol{\xi}_{\rm pop}) \right] \mathcal{L}(\mathcal{D}_{\rm SNe}\mid \boldsymbol{\pi}, \boldsymbol{\xi}_{\rm pop}) \\
    \times p(\boldsymbol{\pi}, \boldsymbol{\xi}_{\rm pop}) d \boldsymbol{\xi}_{\rm pop}.
\end{multline}

Table~\ref{table:param_summary} summarizes the hyper-parameters describing the cosmological parameters, the SNe brightness distribution as well as the lens population that we are sampling hierarchically. We also state the parameter priors we employ in the forecast. We refer to \cite{Birrer:2020} for the formal approximation we are making in the Bayesian analysis while treating other lens model parameters independently among the different lenses and to \cite{Wagner-Carena:2021} to a hierarchical analysis inferring a wider range of lens model hyper-parameters.

The likelihood of an individual lens for a given set of hyper-parameters, $\mathcal{L}(\mathcal{D}^{i}_{\rm L}\mid \boldsymbol{\pi}, \boldsymbol{\xi}_{\rm pop})$, is given by the integral of the individual parameters according to the specified distribution of the hyper-parameters

\begin{equation}\label{eqn:marginal_likelihood_lens}
    \mathcal{L}(\mathcal{D}^{i}_{\rm L}\mid \boldsymbol{\pi}, \boldsymbol{\xi}_{\rm pop}) = \int \mathcal{L}(\mathcal{D}^{i}_{\rm L}\mid \boldsymbol{\pi}, \boldsymbol{\xi}) p(\boldsymbol{\xi} \mid \boldsymbol{\xi}_{\rm pop}) d\boldsymbol{\xi},
\end{equation}
where $p(\boldsymbol{\xi} \mid \boldsymbol{\xi}_{\rm pop})$ is the distribution function of the individual parameters $\boldsymbol{\xi}$ as specified by the population parameters $\boldsymbol{\xi}_{\rm pop}$, and $\mathcal{L}(\mathcal{D}^{i}_{\rm L}\mid \boldsymbol{\pi}, \boldsymbol{\xi})$ is the likelihood specified by Equation~\ref{eqn:lens_likelihood} and its Gaussian form (Eqn. \ref{eqn:gauss_lens_likelihood}) when stating the angular diameter distances as a function of the cosmological parameters $\boldsymbol{\pi}$.
The same statement as for the lens likelihood (Eqn.~\ref{eqn:marginal_likelihood_lens}) applies for the SNe sample likelihood $\mathcal{L}(\mathcal{D}_{\rm SNe}\mid \boldsymbol{\pi}, \boldsymbol{\xi}_{\rm pop})$. The marginalization in $\mathcal{L}(\mathcal{D}_{\rm SNe}\mid \boldsymbol{\pi}, \boldsymbol{\xi}_{\rm pop})$ goes over the supernovae brightness distribution hyper-parameters $\overline{m}_{\rm p}$ and $\sigma(m_{\rm p})$. We note that the SNe distribution parameters are shared for both the SNe population likelihood and the individual lens likelihoods, as well as the cosmological parameters relevant to describe the relative expansion history. The absolute scales of the Universe, stated in the form of $H_0$, only enter explicitly in the time-delay likelihood.

\begin{table*}
\caption{Summary of the model parameters sampled in joint SNe + SL hierarchical inference.}
\begin{center}

\begin{tabular}{l l l}
    \hline
    name & prior & description \\
    \hline \hline
    \multicolumn{3}{|l|}{Cosmology  (Flat $\Lambda$CDM)} \\
    \hline
    $H_0$ [\Hunit] & $\mathcal{U}([0, 150])$  & Hubble constant \\
    $\Omega_{\rm m}$ & $\mathcal{U}([0, 1])$ & current normalized matter density \\
    \hline
    Mass profile \\
    $\overline{\lambda}_{\rm int}$ & $\mathcal{U}([0.5, 1.5])$ & internal MST population mean \\
    $\sigma(\lambda_{\rm int})$ & $=0.03$ & 1-$\sigma$ Gaussian scatter in $\lambda_{\rm int}$ \\ 
    \hline
    SNe population \\
    $\overline{m}_{\rm p}$ & $\mathcal{U}([0, 30])$  & mean of the apparent magnitude distribution of the SNe population at $z_{\rm pivot}=0.1$ \\
    $\sigma(m_{\rm p})$ & $=0.1$  & 1-$\sigma$ Gaussian scatter in intrinsic SNe magnitude distribution at fixed redshift $m_{\rm p}$\\
    \hline
    Line of sight \\
    $\overline{\kappa}_{\rm ext}$ & $=0$  & population mean in external convergence of lenses \\
    $\sigma(\kappa_{\rm ext})$ & $=0.025$  & 1-$\sigma$ Gaussian scatter in $\kappa_{\rm ext}$ \\
    \hline
\end{tabular}

\end{center}
\label{table:param_summary}
\end{table*}

\subsection{Analytic error propagation}\label{sec:analytic_error_propagation}
Before we present the forecast and results with the full hierarchical sampling and propagating of the covariances in the model described in Section~\ref{sec:hierarchical_analysis}, we also provide an analytic, simplified, approximate error propagation. This calculation is easily accessible, fast to compute, and provides valuable insights in the relative importance of different uncertainty components impacting the final $H_0$ constraints.

To first order, the relative $H_0$ uncertainty, $\sigma_{H_0}/H_0$, comprises of the uncertainty in the population mean of the MST parameter\footnote{including internal and external MST effects}, $\overline{\lambda}$, and the uncertainty when performing an uncorrelated error propagation when fixing $\overline{\lambda}={\rm const}$ as
\begin{equation}\label{eqn:h0_error_lambda}
    \frac{\sigma(H_0)}{H_0} \approx \sqrt{ \left(\frac{\sigma(\overline{\lambda})}{\overline{\lambda}}\right)^2 
    + \left(\frac{\sigma(H_0)}{H_0}\right)^2_{\overline{\lambda}={\rm const}}}.
\end{equation}

In the following, we approximate the uncertainty budget for the distinct terms in Equation~\ref{eqn:h0_error_lambda}. For simplicity of this analysis, we assume that for all lenses, and all images, the uncertainty terms are identical. In practice, and in the full inference, inverse uncertainty weighting must be considered.

\subsubsection{Uncertainty terms in the MST}
The first term on the right-hand side of Equation~\ref{eqn:h0_error_lambda} above can be determined with absolute lensing magnifications.
The population level uncertainty in $\overline{\lambda}$ can, to first order, be expressed as the uncertainty in the population mean of the apparent unlensed brightness $\overline{m}_{\rm p}$\footnote{The differential in logarithmic astronomical magnitude $m$ in regards to relative linear flux $I$ is $I \partial m / \partial I = -2.5 \log_{10}(e) \approx -1.086$. Thus small scatter described in astronomical magnitudes are approximately the same scatter in relative flux.}, which is covariant among all lenses, the uncertainty in the relative expansion history translating the apparent magnitude of the distribution of the external SNe sample to the glSNe source redshift, and uncorrelated measurement uncertainties for each individual lens (Eqn.~\ref{eqn:mu_mst_propagation} and \ref{eqn:error_differential}) as

\begin{multline}\label{eqn:lambda_error_propagation}
    \left(\frac{\sigma(\overline{\lambda})}{\overline{\lambda}}\right)^2 \approx 
    \frac{1}{2}\left[\sigma^2(\overline{m}_{\rm p})
    + \sigma^2\left(\frac{L_{z_{\rm source}}}{L_{z_{\rm SNe}}} \right) \right. \\
    + \left. \frac{1}{N_{\rm lens}}\left(\frac{\sigma(\mu_i)}{\mu_i}\right)_{\overline{m}_{\rm p}={\rm const}}^2
    \right],
\end{multline}
where $N_{\rm lens}$ is the number of lens systems.
Furthermore, for simplicity, we assumed equal precision in the individual relative magnification measurements in Equation~\ref{eqn:lambda_error_propagation} above for each lens.

The relative magnification uncertainty per lens with fixed source population mean $\overline{m}_{\rm p}$, can be written following Equation~\ref{eqn:error_differential} as

\begin{multline}\label{eqn:magnification_error_propagation}
 \left(\frac{\sigma(\mu_i)}{\mu_i}\right)_{\overline{m}_{\rm p}={\rm const}}^2 \approx 
\left(\frac{\sigma(F_{\rm unl})}{F_{\rm unl}}\right)^2 +
 \frac{1}{4}\left(\frac{\sigma(F_{\rm obs})}{F_{\rm obs}} \right)^2\\
 + \left(\frac{\sigma(\mu_{\rm macro})}{\mu_{\rm macro}} \right)^2
 + \frac{1}{4}\left(\frac{\sigma(\Delta\mu_{\rm milli})}{\Delta\mu_{\rm milli}} \right)^2
 + \frac{1}{4}\left(\frac{\sigma(\Delta\mu_{\rm micro})}{\Delta\mu_{\rm micro}} \right)^2.
\end{multline}
The first term on the right hand side of the equation above is the intrinsic scatter in the standardizable source, the second is the flux measurement uncertainty, and the following ones are the different scales of the lensing effect. The factor $1/4$ comes from the fact that we consider quadruply lensed quasars as this approximation assumes the random errors in the milli- and micro-lensing effects to be uncorrelated among the different images. The macro-model magnification uncertainties are covariant and thus we omit the factor $1/4$ in the approximation.

\subsubsection{Time-delay and Fermat potential uncertainties}
The second term on the right hand side of Equation~\ref{eqn:h0_error_lambda} encompasses all other sources of uncertainties not related to global inference shifts due to the MST.
In particular, this involves uncertainties in the time-delay measurements, the Fermat potential uncertainty for a specified mass profile family (in our case PEMD+shear) from high-resolution imaging data, and the random uncertainties in the LOS convergence estimates and the internal MST.
In addition, we include in this second term uncertainties in the relative expansion history that translate the angular diameter distance measurements to the lensing system, $D_{\Delta t}$, relative to to the scales at current time, and thus $H_0$, which we denote as $\sigma(H_0/ D_{z={\rm SL}})$.

In terms of distance measurements, we can approximately write

\begin{multline}\label{eqn:error_lambda_fixed}
    \left(\frac{\sigma(H_0)}{H_0}\right)^2_{\overline{\lambda}={\rm const}} \approx
    \frac{1}{N_{\rm lens}} \left(
    \left(\frac{\sigma\left(D^{\rm pl}_{\Delta t}\right)}{D^{\rm pl}_{\Delta t}}\right)^2
    + \left(\frac{\sigma(\lambda)}{\lambda}\right)^2 \right) \\
    + \sigma^2\left(\frac{H_0}{D_{z={\rm SL}}}\right),
\end{multline}
where the relative time-delay distance measurement uncertainty can be estimated by the relative Fermat potential uncertainties from imaging modeling and the relative time-delay uncertainties
\begin{equation}\label{eqn:ddt_gaussian_error}
    \left(\frac{\sigma\left(D^{\rm pl}_{\Delta t}\right)}{D^{\rm pl}_{\Delta t}}\right)^2 \approx \left(\frac{\sigma(\Delta\tau_{\rm pl})}{\Delta\tau_{\rm pl}}\right)^2
    + \left(\frac{\sigma(\Delta t)}{\Delta t}\right)^2,
\end{equation}
and the scatter and random uncertainty in $\lambda$ coming from the internal and external scatter, which can be approximated as
\begin{equation}
    \left(\frac{\sigma(\lambda)}{\lambda}\right)^2 =
    \left(\frac{\sigma(\lambda_{\rm int})}{\lambda_{\rm int}}\right)^2 + \left(\frac{\sigma(\kappa_{\rm ext})}{1-\kappa_{\rm ext}}\right)^2.
\end{equation}
The time-delay distance uncertainty per lens (Eqn.~\ref{eqn:ddt_gaussian_error}) is, to first order, a weighted product of all the different images. The random uncertainty in the MST acts as a noise term for the individual distance measurements for each lens.

\section{Forecast} \label{sec:forecasts}
Having formulated the methodology and parameterization in the previous sections, we perform different forecast scenarios based on predicted number of glSNe, quality of measurements and systematics effect.
In Section~\ref{sec:lens_population} we state the expected number of glSNe and time-delay measurements and our assumptions on milli- and micro-lensing effects in the magnification. In Section~\ref{sec:lens_uncertainty} we state the lens model, source configuration and uncertainties expected from imaging data on the Fermat potential and magnifications. In Section~\ref{sec:sne_field_scenario} we present the scenario for current and future unlensed SNe data sets. Finally, in Section~\ref{sec:forecast_results} we present the inference results for the different forecast scenarios.

\subsection{Lens population, time-delay and magnification uncertainties}\label{sec:lens_population}
In this work, we focus on the discoveries expected by the Vera Rubin Observatory Legacy Survey of Space and Time (LSST). We do not perform an independent forecast and derive our fiducial forecast scenario based on previous work in the literature.

\subsubsection{Expected glSNe with LSST}\label{sec:glSNe_number}

\citet{Goldstein:2017, Goldstein:2018} estimated, based on the catalogue by \cite{OM10}, the number of glSNe~Ia to be up to 500-900 in 10 years of LSST with unresolved photometric magnification detection where the brightest SN image reaches a peak apparent i-band magnitude of 22.15 or brighter.
\citet{Wojtak:2019} compared two different discovery techniques, by magnifaction and resolved image multiplicity and estimated the annual discovery rate with LSST to be 61 with magnification, 44 with resolved image multiplicity and 89 in hybrid discovery scheme.

It has been noted that lensed supernovae found via image multiplicity exhibit longer time delays and larger image separations making them more suitable for cosmological constraints than their counterparts found via magnification \citep{Wojtak:2019, Huber:2019}.
\citet{Huber:2019} finds, when restricting the expected time-delay measurement to a minimum precision of $<5\%$ and an accuracy of $<1\%$ (if based solely on LSST observations) would reduce the number of lensed type Ia supernovae to about 1 per year. This rate can be increased by a factor of 2–16 by employing other instruments for follow-up observations.
Beyond LSST, for example, \cite{Pierel:2021} predicts that the \textit{Roman} observatory will discovery $\sim 11$ glSNe~Ia.
With follow up efforts in measuring the time delays of the sub-sample restricted on the most promising time-delay measurements \citep{Huber:2019}, LSST+follow up is able to provide $< 1\%$ overall statistical precision on $H_0$ form the time-delay uncertainties of 20 glSNe~Ia \citep[e.g.,][]{Suyu:2020}.

\subsubsection{Milli- and micro-lensing}\label{sec:micro_lensing}
Milli- and micro-lensing effects on the magnification of the glSNe can significantly impact the ability of glSNe to be used as standardizable candles.
Milli-lensing, an effect caused by dark subhaloes of the main deflector or along the line of sight \citep[e.g.,]{Dalal:2002, Gilman:2020wdm, Hsueh:2020}, or baryonic effects \citep[e.g.,][]{Hsueh:2016, Gilman:2017}.
Flux ratio anomalies at the $\sim 10\%$ level have been studied and used to constrain dark matter properties with quardruply lensed quasar flux ratio anomalies \citep[e.g.,][]{Gilman:2020wdm, Hsueh:2020}.
For physical source size of SNe, Kelly et al. (in prep) estimated for SN Refsdal \citep{Kelly:2015} about a $\sim 10\%$ scatter from milli-lensing based on the forward modeling methodology by \cite{Gilman:2019, Gilman:2020wdm}.

Microlensing caused by stars or other compact objects in the foreground lens- ing galaxy or along the line of sight can be a more significant limit to the standardization of glSNe. Microlensing can independently magnify or de-magnify individual images of the background source \citep{Dobler:2006, Bagherpour:2006}, introducing scatter into the shape and amplitude of the resulting light curves. The effect of microlensing on each lensed image depends on the local smooth lensing properties (convergence $\kappa$, shear $\gamma$) and the stellar (or compact) projected mass fraction $\kappa_{*}/\kappa$.
For example, \cite{Schechter:2002} investigated stellar micro-lensing effects on lensed quasars at image magnifications of $\mu\sim 10$ with moderate compact object mass fractions and showed that for such scenarios, the expected micro-lensing scatter can result in more than an astronomical magnitude.

\cite{Foxley-Marrable:2018}, with the aim of assessing glSNe~Ia to be standardizable in the same spirit as this work, evaluated the effect of microlensing on glSNe~Ia for various image configurations.
They found that there are regions of parameter space where the effect of microlensing is suppressed enough for the glSN~Ia to be standardizable. Specifically, regions of low $\kappa$, $\gamma$ and high $s$ are subject to microlensing scatter of $\sigma_{\rm ML} \sim 0.15$ in astronomical magnitude, particularly at early times. Physically, this corresponds to asymmetric configurations with at least one image located far outside the Einstein radius, which will experience the least amount of microlensing.

When \cite{Foxley-Marrable:2018} combined their microlensing models with the glSNe~Ia catalogue from \cite{Goldstein:2017}, they predicted that $\sim 22\%$ of the $\sim 930$ glSNe~Ia to be discovered by LSST will be standardizable ($\sigma_{\rm ML} \sim 0.15$ or below for at least one image). The standardizabe sample has a median maximal time delay of 44 days and consists of 5:1 ratio of doubles vs quads.
\cite{Foxley-Marrable:2018} further concluded that from their sample of 650 glSNe~Ia, of which accurate time delays can be measured, the MSD can be broken at the 0.5\% level when considering microlensing and intrinsic scatter of the SNe as the source of statistical uncertainties.

\subsubsection{Specific numbers and uncertainties of this forecast}

Overall, restricting the follow-up effort to a considerably smaller number than the overall expected discoveries optimized to derived time-delay precision and accuracy, LSST is expected to provide sufficient statistical precision on time delays with a sub-percent error budget on final $H_0$ constraints.
However, using glSNe for standardizable magnification constraints may require a larger and potentially different subset of the glSNe~Ia population to be further investigated with follow-up efforts. Given the mass profile uncertainties are at the 10\% level for individual lenses, we consider in this forecast a scenario with an extended sample of glSNe~Ia beyond the subset of \cite{Huber:2019, Suyu:2020} with lower precision time-delay measurements, including both glSNe with shorter time delays as well as fainter images.

In this forecast, we design a scenario where time-delay precision and standardizable nature of glSNe~Ia can be utilized. We stress that time-delay measurement and flux standardization do not necessary need to come from the same lenses\footnote{The lenses need to be self-similar to translate the MST-breaking to the time-delay lenses}.

We chose a lens population roughly following \cite{Foxley-Marrable:2018}.
In total, we perform our forecast with 144 glSNe, among which 24 are quads and 120 are doubles. For the quad population, we split the sample in 8 crosses, 8 cusps and 8 fold configurations\footnote{This split is not based on ability of standardizable magnifications, but primarily for pedagogic illustration.}. The doubles we split into three different configurations each consisting of 40 systems.

For the redshift distribution, we assume a uniform distribution of the deflector redshift, $z_{\rm lens}$, between $z=0.1$ and $z=0.5$, and for the source redshift, $z_{\rm source}$, a uniform distribution in $\mathcal{U}[z_{\rm lens}+0.2, 1.]$, similar as the distribution by \cite{Huber:2019, Suyu:2020} restricting to the brighter population for both accurate time-delay and flux measurements.
We stress the importance of rapid spectroscopic follow-up to confirm the SNe type and we assume that the follow-up has been acquired for the SN sample and the SNe have been robustly typed.

For the time-delay measurement, we assume that the light curves can be resolved in follow up observations and the relative time delays can be measured with a precision of 2 day per image pair\footnote{We refer to Equation~\ref{eqn:ddt_gaussian_error} for the impact on the statistical error propagation with different time-delay precision. Overall, in this forecast the time-delay measurement uncertainties are subdominant to other sources of uncertainties.}. Along with spectroscopy obtained for the typing, these cadenced observations provide further evidence to distinguish the normal SNe~Ia from peculiar subtypes \citep[e.g., see][for review]{Taubenberger:2017}, since fast-declining and super-Chandra subtypes do not show a second maximum in the NIR, unlike normal SNe~Ia. The presence of an NIR second maximum was further confirmation that iPTF16geu is a normal SN~Ia \citep{Dhawan:2020}.
In addition to precise time delays, obtaining resolved photometry, in multiple wavebands, is crucial for constraining the extinction properties. We assume that similar to the case for iPTF16geu, there are cadenced observations in multiple optical and NIR filters to constrain the extinction in the host galaxy and the individual lines of sight in the lens for each image \citep{Dhawan:2020}. Accounting for extinction correction, the magnification is inferred robustly with small uncertainties, making it a subdominant contribution to other sources.

For the flux uncertainty at peak brightness for the individual images, we use an effective relative magnitude uncertainty $\sigma_{\rm eff}(m)$ that includes possible uncertainties from small scale milli- and micro-lensing effects

\begin{multline}\label{eqn:F_eff}
    \sigma^2_{\rm eff}(m) \equiv \left(\frac{\sigma(F_{\rm obs})}{F_{\rm obs}} \right)^2
 + \left(\frac{\sigma(\Delta\mu_{\rm milli})}{\Delta\mu_{\rm milli}} \right)^2\\
 + \left(\frac{\sigma(\Delta\mu_{\rm micro})}{\Delta\mu_{\rm micro}} \right)^2.
\end{multline}
This is a practically convenient noise definition when assuming Gaussian uncorrelated error in terms of the uncertainty relevant to constraining the MST.
Beyond the intrinsic scatter in the SNe population, $\sigma(m_{\rm p})$, and the uncertainty in the macro-model magnification, $\sigma(\mu_{\rm macro})$, the term in Equation~\ref{eqn:F_eff} above can play a dominant role in the uncertainty budget and is by itself uncertain given the current rare discoveries and follow-up data of glSNe systems.

We separate the $\sigma_{\rm eff}(m)$ term for the different images into one image denoted as the standardizable one, $\sigma_{\rm eff, std}(m)$, and all other images denotedas the microlensing dominated ones, $\sigma_{\rm eff, ML}(m)$.
For $\sigma_{\rm eff, ML}(m)$ we assume a scatter of one magnitude, $\sigma_{\rm eff, ML}(m)=1.0$, making most images of glSNe inefficient probes of the mass profile.

For the 'standardizable image' we perform three different scenarios for $\sigma_{\rm eff, std}(m)$. The first scenario, denoted as $\textsc{ideal}$, sets $\sigma_{\rm eff, std}(m) = 0$ for all measurements, assuming no milli- and micro-lensing effects and perfect flux measurements. The $\textsc{ideal}$ scenario is meant to assess the error budget and the precision floor of any other uncertainty component.
The second scenario, denoted as $\textsc{realistic}$, sets $\sigma_{\rm eff, std}(m) = 0.2$ for all measurements. The $\textsc{realistic}$ scenario represents a likely scenario for the uncertainty terms contained in $\sigma_{\rm eff, std}(m)$. A specific split among its constituents is not required but is motivated by a $<10\%$ flux measurement uncertainty, a $\sim 10\%$ milli-lensing uncertainty, and a $\sim 15\%$ micro-lensing uncertainty.
The third scenario, denoted as $\textsc{extreme}$, sets $\sigma_{\rm eff, std}(m) = 1$, a scenario where the magnification of every single image of a glSNe is dominated by small scale micro-lensing magnification.
We highlight, that these uncertainty terms should be interpreted as statistical averages for the population of glSNe. In particular, the micro-lensing component is expected to vary from image to image substantially depending on the stellar surface brightness.

Table~\ref{table:glSNe_population} summarizes our choices for the forecasts presented in this work. We emphasize that our forecast scenario and numbers operate under the assumption of imminent and complete follow-up observation after a discovery or promising candidate. The total number and numbers per year may be lower when the dedicated follow-up and we provide an extended forecast prediction as a function of glSNe in Section~\ref{sec:error_budget}.

\begin{table*}
\caption{glSNe forecast scenarios in terms of numbers of glSNe, their redshift distribution and measurement uncertainties. The parameters of the macro model, and their uncertainties for the forecast, are presented in Table~\ref{table:lens_param}. The effective magnitude precision (Eqn. \ref{eqn:F_eff}) is split between one image that is less affected by microlensing ($\sigma_{\rm eff, std}$) and to the other images more strongly affected by microlensing ($\sigma_{\rm eff, ML}$).}
\begin{center}
\begin{tabular}{l l l}
    \hline\hline
    \multicolumn{3}{|l|}{Number of glSNe}\\
    \hline
    cusp & 8 \\
    cross & 8 \\
    fold & 8 \\
    doubles & 40 + 40 + 40\\
    \hline
    total & 144 \\
    \hline \hline
    \multicolumn{3}{|l|}{Redshift distribution} \\
    \hline
    $z_{\rm lens} $ & $\mathcal{U}[0.2, 0.5]$ & deflector redshift \\
    $z_{\rm source} $ & $\mathcal{U}[z_{\rm lens}+0.2, 1.0]$ & source redshift \\
    \hline \hline
    \multicolumn{3}{|l|}{Measurement uncertainties (1-sigma)} \\
    \hline
    $\delta\Delta t$ & $\pm 2.0$ days & time-delay precision \\
    \hline
    $\sigma_{\rm eff}$ & Effective magnitude precision (Eqn. \ref{eqn:F_eff}) & scenario \\
    \hline
    $\sigma_{\rm eff, std}$ & $\pm 0.0$ & \textsc{ideal}\\
    $\sigma_{\rm eff, ML}$ & $\pm 1.0$ \\
    \hline
    $\sigma_{\rm eff, std}$ & $\pm 0.2$ & \textsc{realistic}\\
    $\sigma_{\rm eff, ML}$ & $\pm 1.0$ \\
    \hline
    $\sigma_{\rm eff, std}$ & $\pm 1.0$ & \textsc{extreme}\\
    $\sigma_{\rm eff, ML}$ & $\pm 1.0$ \\
    \hline
\end{tabular}
\end{center}
\label{table:glSNe_population}
\end{table*}

\subsection{Deflector model}\label{sec:lens_uncertainty}

The model parameters for the PEMD+shear model are described in Table~\ref{table:lens_param}. We chose the same lens model for all glSNe systems for simplicity, but with general application of the error propagation and uncertainties. The different source position of the glSNe for the cusp, cross, and fold configurations are also provided in Table~\ref{table:lens_param}.
We mimic high-resolution imaging data constraints on the lens model parameters with Gaussian errors on the lens model parameters, also stated in Table~\ref{table:lens_param}. For the image positions of the multiply lensed SNe, we assume an astrometric precision of $\pm 0.005$ arcseconds, achievable with high-resolution imaging around SNe peak brightness. 
\cite{BirrerTreu:2019} highlighted the importance and requirements on the astrometric precision of the images of the time-variable sources. Our chosen precision meets the requirement not to be the dominant uncertainty in our inference.

We sample the posterior of the imaging data (Eqn.~\ref{eqn:imaging_likelihood}) with the Gaussian likelihood in the lens model and image position parameters while demanding the image positions originating from the same source position for the proposed lens model as a solution of the lens equation.
We then transform the posteriors into the relative Fermat potential and absolute magnifications at the predicted image positions (Eqn.~\ref{eqn:imaging_likelihood}).

\begin{figure*}
  \centering
  \includegraphics[angle=0, width=\textwidth]{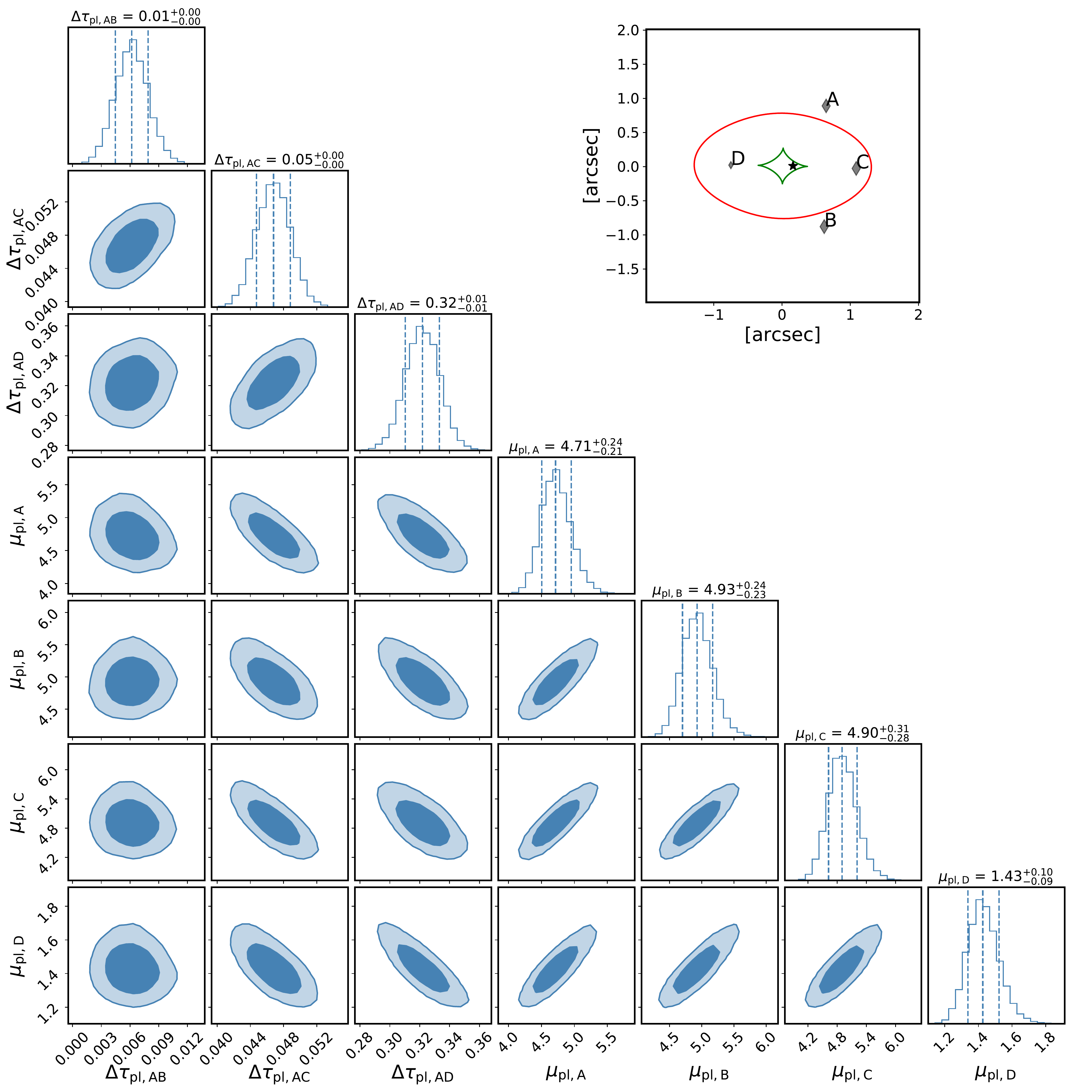}
  \caption{Mock image modeling posterior on the relative Fermat potential and lensing magnification between the image positions of a glSNe when fit by a PEMD+shear lens model for the cusp configuration. The lens model parameters and uncertainties are presented in Table~\ref{table:lens_param}. The configuration of the image position (diamonds), inner caustic (green) and critical curve (red) are illustrated in the top right figure. The posteriors for the cross and fold configurations are presented in Appendix~\ref{app:imaging_posteriors}.  \href{\notebooklink}{\faGithub}} %
\label{fig:cusp_imaging_posteriors}
\end{figure*}

The joint relative Fermat potential and magnification posteriors for the cusp configuration are illustrated in Figure~\ref{fig:cusp_imaging_posteriors}.
Similar posterior products are derived for the cross and fold configurations and are presented in the Appendix~\ref{app:imaging_posteriors} (Figs.~\ref{fig:cross_imaging_posteriors}, \ref{fig:fold_imaging_posteriors}).

The effective macro-model magnification uncertainty is $\sim$ 5\% per image. The effective relative Fermat potential uncertainty is $\sim$ 4\% per image pair. The uncertainties are comparable for the three different image configurations chosen in this forecast and compatible with uncertainties obtained from the analysis of real data by the H0LiCOW/SHARP/STRIDES/TDCOSMO collaborations \citep[][]{Suyu:2010, Suyu:2013, Wong:2017, Birrer:2019, Chen:2019, Rusu:2020, Shajib:2020strides}.
The posteriors in Fermat potential and magnification for our chosen configurations and uncertainties are well approximated by multivariate Gaussians, which justifies the use the Gaussian likeihood of Equation~\ref{eqn:imaging_likelihood_gaussian} with the covariance matrix $\mathbf{\Sigma}_{\Delta\tau\mu}$.

\begin{table}
\caption{Deflector model parameters and uncertainties for the forecast. The parameters correspond to the PEMD+shear model. The uncertainties represent high resolution imaging data modeling uncertainties when the SNe is faded away.}
\begin{center}
\begin{tabular}{l r l}
    \hline
    Parameter & value & description \\
    \hline \hline
    \multicolumn{3}{|l|}{Lens model (PEMD+shear)}\\
    \hline
    $\theta_{\rm E}$ & $1.0\pm 0.02$ & Einstein radius [arcsec]\\
    $\gamma_{\rm pl}$ & $2.00 \pm 0.03$ & Power-law slope\\
    $e_1$ & $0.30\pm0.01$ & Eccentricity of deflector\\
    $e_2$ & $-0.01\pm0.01$ & Eccentricity of deflector\\
    $x_0$ & $0.00\pm0.01$ & Center of deflector [arcsec]\\
    $y_0$ & $0.00\pm0.01$ & Center of deflector [arcsec]\\
    $\gamma_1$ & $0.05\pm0.01$ & External shear component\\
    $\gamma_2$ & $0.00\pm0.01$ & External shear component\\
    \hline
    \multicolumn{3}{|l|}{Image configurations} \\
    \hline
    $(x_{\rm s}, y_{\rm s})$ & $(0.15, 0.00)$ & cusp source position [arcsec]\\
    $(x_{\rm s}, y_{\rm s})$ & $(0.02, 0.00)$ & cross source position [arcsec]\\
    $(x_{\rm s}, y_{\rm s})$ & $(0.05, 0.10)$ & fold source position [arcsec]\\
    $(x_{\rm s}, y_{\rm s})$ & $(0.05, 0.40)$ & double\#1 source position [arcsec]\\
    $(x_{\rm s}, y_{\rm s})$ & $(0.20, 0.20)$ & double\#2 source position [arcsec]\\
    $(x_{\rm s}, y_{\rm s})$ & $(0.40, 0.20)$ & double\#3 source position [arcsec]\\
    \hline
    $\delta\theta_{\rm image}$ & $\pm 0.005$ & astrometric precision [arcsec]\\
    \hline
\end{tabular}
\end{center}
\label{table:lens_param}
\end{table}

\subsection{Unlensed field SNe data set}\label{sec:sne_field_scenario}

The data set of unlensed (field) SNe fulfils two purposes. First, it anchors the apparent unlensed population of SNe, $\overline{m}_{\rm p}$ and $\sigma(m_{\rm p})$, and their uncertainties. The parameter $\overline{m}_{\rm p}$ directly translates to $\overline{\lambda}$, and thus to $H_0$.
Second, the relative luminosity distances of SNe constrain the relative expansion history of the Universe, and thus $\Omega_{\rm m}$ in flat $\Lambda$CDM. The uncertainty on the relative expansion history can have two ways to impact the resulting $H_0$ uncertainty: 
(i) the translation of the distance measurement corresponding to the glSNe systems at intermediate redshifts to the local distance constraints for a given MST parameter $\overline{\lambda}$ (Eqn.~\ref{eqn:error_lambda_fixed}), similar to an inverse distance ladder; 
(ii) the translation of the apparent magnitudes from the distribution of unlsensed (mostly lower redshifts) to the glSNe source redshifts (mostly higher redshifts) (Eqn.~\ref{eqn:lambda_error_propagation}). 

To assess current and future uncertainties coming from field SNe data sets, we set up two scenarios. First, we utilize the Pantheon data set \citep{Scolnic:2018}. In particular, we are using the full covariance matrix product as described by \citet{Scolnic:2018}. The covariance matrix includes the intrinsic scatter in the SN~Ia distribution as well as covariant systematic uncertainties.
Second, we mimic a future SNe data set with an anticipated increase in the sample and lowering of systematics over the coming 10 years with the onset of the \textit{Roman Space Telescope}. We use the forecast covariance matrix by \cite{Hounsell:2018}.
The comparison between the hierarchical glSNe inference with the current Pantheon sample and the future SNe sample allows us to emphasize the importance of the field SNe sample in the next decade to utilizing glSNe to their full potential.

Table~\ref{table:sne_samples} provides the one dimensional marginal constraints on $\Omega_{\rm m}$ and $\overline{m}_{\rm p}$ derived from the two samples.

\begin{table}
\caption{Summary of constraints provided by the two field SNe samples used in the forecast, the \emph{Pantheon} sample by \citet{Scolnic:2018}, and a forecast for the \emph{Roman Space Telescope} by \citet{Hounsell:2018}.}
\begin{center}
\begin{tabular}{l l l l}
\hline
Scenario & $\Omega_{\rm m}$ & $\overline{m}_{\rm p}$ & $\sigma(m_{\rm p})$  \\ 
\hline \hline
\textsc{Pantheon} & ${0.299}_{-0.022}^{+0.023}$ & ${18.966}_{-0.008}^{+0.008}$ & $=0.1$ \\ 
\textsc{Roman} & ${0.300}_{-0.004}^{+0.005}$ & ${18.966}_{-0.004}^{+0.005}$ & $=0.1$ \\ 
\hline

\end{tabular}
\end{center}
\label{table:sne_samples}
\end{table}

\subsection{Forecast results}\label{sec:forecast_results}

\begin{figure*}
  \centering
  \includegraphics[angle=0, width=130mm]{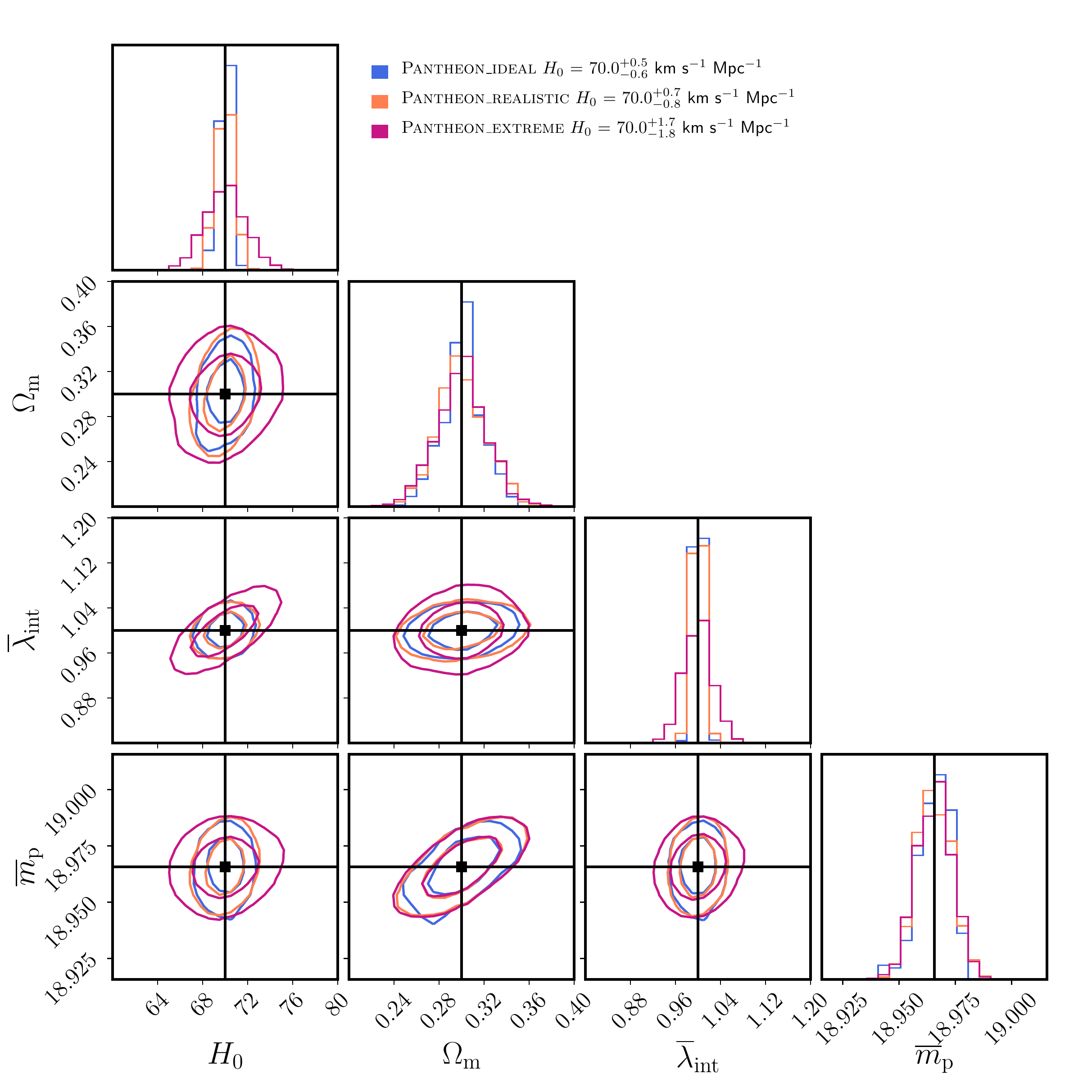}
  \caption{Posterior inference for the forecast of 144 glSNe of the parameters and their priors presented in Table~\ref{table:param_summary} (see also  Tables~\ref{table:glSNe_population} and \ref{table:lens_param} for details on the uncertainties) with the Pantheon unlensed SNe sample.
  We specified three different uncertainty scenarios for $\sigma_{\rm eff, std}(m)$ (Eqn.~\ref{eqn:F_eff}, Section~\ref{sec:glSNe_number}), \textsc{ideal} (blue; $0.0$), \textsc{realistic} (orange; $0.2$), and \textsc{extreme} (violet; $1.0$).
  Figure~\ref{fig:forecast_posteriors_future} presents the same forecast with a Roman unlensed SNe sample.
  \href{\notebooklink}{\faGithub}} %
\label{fig:forecast_posteriors_pantheon}
\end{figure*}

\begin{figure*}
  \centering
  \includegraphics[angle=0, width=130mm]{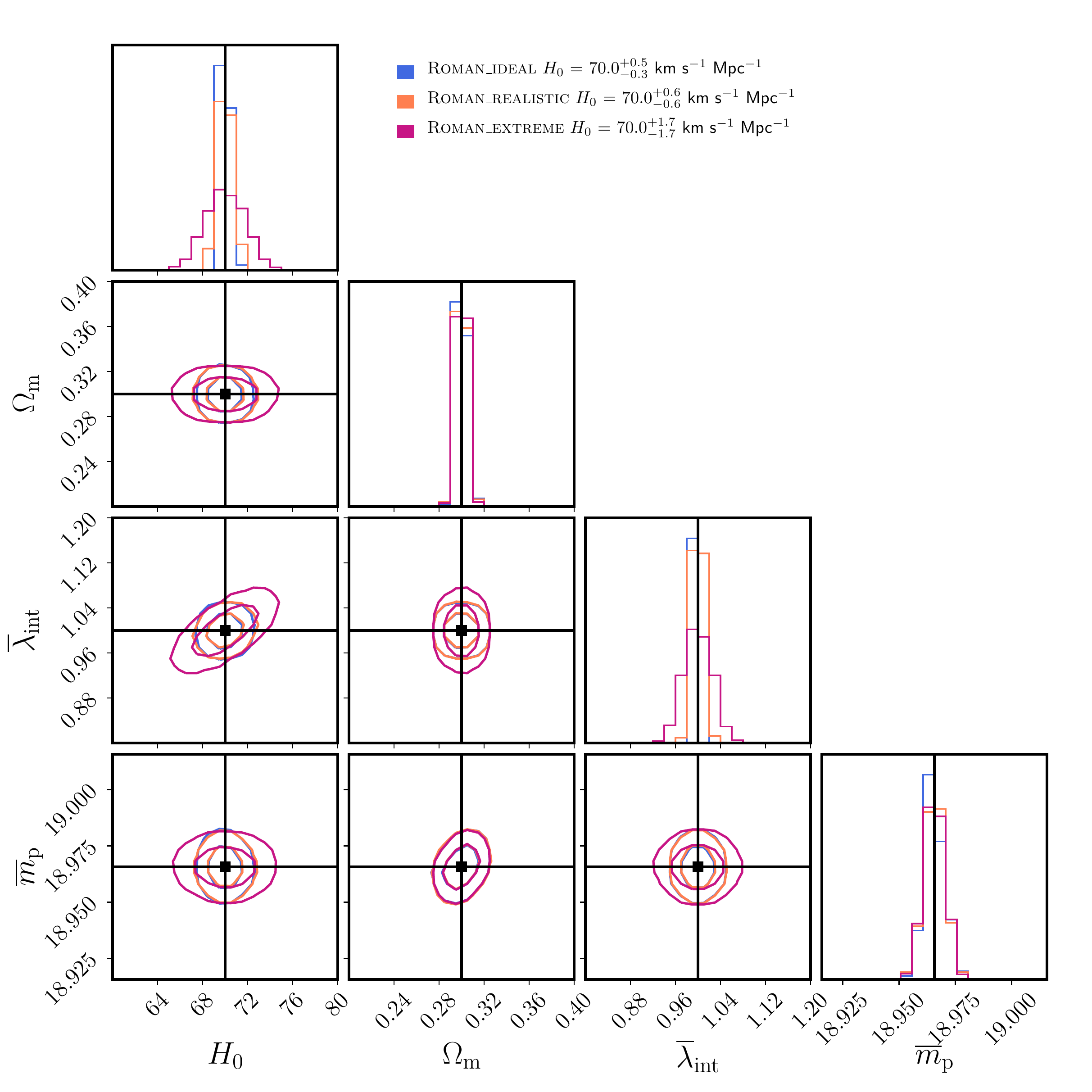}
  \caption{
  Posterior inference for the forecast of 144 glSNe of the parameters and their priors presented in Table~\ref{table:param_summary} (see also  Tables~\ref{table:glSNe_population} and \ref{table:lens_param} for details on the uncertainties) with a Roman unlensed SNe sample (Table~\ref{table:sne_samples}).
  We specified three different uncertainty scenarios for $\sigma_{\rm eff, std}(m)$ (Eqn.~\ref{eqn:F_eff}, Section~\ref{sec:glSNe_number}), \textsc{ideal} (blue; $0.0$), \textsc{realistic} (orange; $0.2$), and \textsc{extreme} (violet; $1.0$).
  Forecast
  Figure~\ref{fig:forecast_posteriors_pantheon} presents the same forecast with the current Pantheon unlensed SNe sample.\href{\notebooklink}{\faGithub}} %
\label{fig:forecast_posteriors_future}
\end{figure*}

We perform the hierarchical analysis of the parameters and their priors presented in Table~\ref{table:param_summary}. We make use of the Gaussian likelihoods of individual glSNe system as presented in Section~\ref{sec:method} with the numbers of glSNe and uncertainties presented in Tables~\ref{table:glSNe_population} and \ref{table:lens_param}.
We specified three different uncertainty scenarios for $\sigma_{\rm eff, std}(m)$ (Eqn.~\ref{eqn:F_eff}, Section~\ref{sec:glSNe_number}), \textsc{ideal} ($0.0$), \textsc{realistic} ($0.2$), and \textsc{extreme} ($1.0$). We also specified two different unlensed SNe scenarios, \textsc{Pantheon} and \textsc{Future} (Table~\ref{table:sne_samples}). Any combination of SNe sample and $\sigma_{\rm eff, std}(m)$ uncertainties results in six forecast scenarios.
Figure~\ref{fig:forecast_posteriors_pantheon} shows the posterior inference with the scenarios of the \textsc{Pantheon} sample. Figure~\ref{fig:forecast_posteriors_future} shows the same inferences with the \textsc{Roman} sample.

In addition to these six inferences with a fully covariant MST component in the deflector model, we perform, as a reference for the time-delay and PEMD+shear lens model uncertainties, the forecast also without a covariant MST component by fixing $\overline{\lambda}_{\rm int}=1$ for both SNe scenarios. The scenarios without the MST do not depend on the error budget of the lensing magnifications $\sigma_{\rm eff}(m)$ and the difference in the unlensed SNe sample and the glSNe sample only impacts the translation of the distance measurements into $H_0$.
Table~\ref{tab:results} summarizes the results in regard of the relative precision on $H_0$ for the eight different scenarios considered in this work.

First, ignoring the MST, the mock data of measured time delays and Fermat potential allow one to constrain $H_0$ to 0.5\% precision with both, \textsc{Pantheon} and \textsc{Roman} sample. This set of forecast serves as a statistical reference and do not require standardizable magnifications to add information.

Once the MST is let free and only constrained by the magnifications, both the impact of the uncertainties of $\sigma_{\rm eff}{m}$ and the external SNe sample significantly impact the resulting constraints. The difference between the constraining power of the \textsc{Pantheon} and \textsc{Roman} sample can be seen prominently when comparing the scenarios with $\sigma_{\rm eff, ml}{m}=0$, the \textsc{ideal} case without micro-lensing. The \textsc{Pantheon} inference results in a precision of 0.8\% while the increased constraining power of the \textsc{Roman} sample results in a 0.6\% precision on $H_0$. The error budget of the \textsc{Pantheon\_ideal} scenario is dominated by uncertainties in the unlensed SNe population whereas the \textsc{Roman\_ideal} achieves almost the same precision as a scenario without an MST uncertainty. 

When including \textsc{realistic} or even \textsc{extreme} micro-lensing uncertainties in our forecast, the uncertainties in $\overline{\lambda}_{\rm int}$ start dominating the constraining power on $H_0$ as expected from the constraining power on the magnification constraints (Eqns~\ref{eqn:lambda_error_propagation}, \ref{eqn:magnification_error_propagation}).
Overall, we highlight our fiducial future scenario, \textsc{Roman\_realistic}, which provides a 0.9\% precision measurement on $H_0$ with a full 10-years LSST survey paired with a \textsc{Roman} supernovae sample.

\begin{table*}
    \centering
    \begin{tabular}{l l  r r r }
\hline
Scenario &  SNe sample  & $\sigma_{\rm eff, std}(m)$ & $\sigma_{\rm eff, ml}(m)$ & $\delta H_0/H_0$  \\ 
\hline
\hline
 \textsc{Pantheon\_no\_mst} & Pantheon  & - & - &  0.6\%  \\ 
 \textsc{Pantheon\_ideal} & Pantheon  & 0.0 & 1.0 & 0.8\%  \\ 
 \textsc{Pantheon\_realistic} & Pantheon  & 0.2 & 1.0  & 1.1\%  \\ 
 \textsc{Pantheon\_extreme} & Pantheon  & 1.0 & 1.0  & 2.5\%  \\ 
 \textsc{Roman\_no\_mst} & Roman SNe & - & - & 0.6\%  \\ 
 \textsc{Roman\_ideal} & Roman SNe & 0.0 & 1.0  & 0.6\%  \\ 
 \textsc{Roman\_realistic} & Roman SNe & 0.2 & 1.0  & 0.9\%  \\ 
 \textsc{Roman\_extreme} & Roman SNe & 1.0 & 1.0  & 2.4\%  \\ 
\hline
    \end{tabular}
    \caption{Summary of the achieved precision on $H_0$ for the six forecast scenarios of this work, and the two scenarios when keeping $\overline{\lambda}$ fixed. We specified three different uncertainty scenarios for the standardizable image $\sigma_{\rm eff, std}(m)$ (Eqn.~\ref{eqn:F_eff}, Section~\ref{sec:glSNe_number}), \textsc{ideal} ($0.0$), \textsc{realistic} ($0.2$), and \textsc{extreme} ($1.0$). We also specified two different unlensed SNe scenarios, \textsc{Pantheon} and \textsc{Future} (Table~\ref{table:sne_samples}). Any combination of SNe sample and $\sigma_{\rm eff}(m)$ uncertainties results in six forecast scenarios. The resulting posterior inference on $H_0$ are given in the last row. The posteriors are also presented in Figure~\ref{fig:forecast_posteriors_pantheon} for the \textsc{Pantheon} and Figure~\ref{fig:forecast_posteriors_future} for the \textsc{Future} supernova sample, respectively.}
    \label{tab:results}
\end{table*}

\subsection{Generalized forecast and expected timeline} \label{sec:error_budget}
Overall, the results of the full hierarchical inference performed in Section~\ref{sec:forecast_results} can be well approximated with the analytical error propagation terms of Section~\ref{sec:analytic_error_propagation}. In this section, we make use of the analytic error propagation and generalize the forecast results of Section~\ref{sec:forecast_results} for a range in the number of glSNe.

\begin{figure*}
  \centering
  \includegraphics[angle=0, width=130mm]{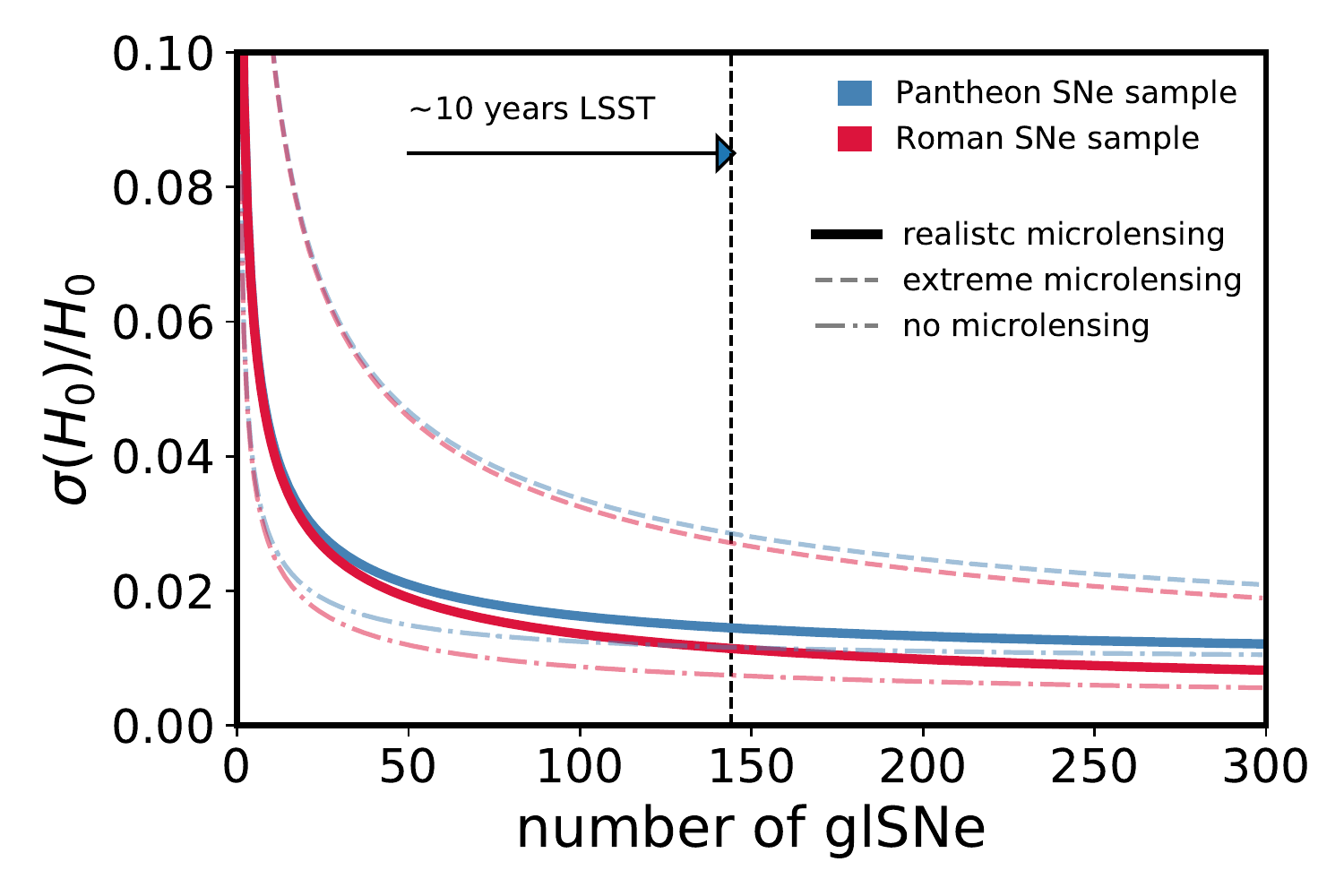}
  \caption{Expected relative precision on $H_0$ as a function of the number of gravitationally lensed SNe (glSNe) with one standardizable image magnification. Blue curves indicate the forecast with the Pantheon SNe sample \citep{Scolnic:2018} and red curves with a future Roman SNe sample \citep{Hounsell:2018}. Thick solid lines mark our \textsc{realistic} expectations of micro-lensing (relative uncertainty of $\sigma_{\rm eff, std}(m)$ (Eqn.~\ref{eqn:F_eff}) of 0.2). Dashed-dotted lines mark the \textsc{ideal} scenario of zero micro-lensing (relative uncertainty of $\sigma_{\rm eff, std}(m)$ of 0.0). Dashed lines mark an extreme micro-lensing scenario (relative uncertainty of $\sigma_{\rm eff, std}(m)$ of 1.0).
  The vertical black dashed line marks the expected number of 144 glSNe for a 10 year LSST survey with one image being only marginally affected by microlensing \citep{Goldstein:2017, Foxley-Marrable:2018} and an assumed optimal follow-up effort enabling the analysis.
  The quality and systematic uncertainties in the unlensed field SNe sample significantly impacts the uncertainty budget for the expected number of glSNe.
  Calculations are made with the analytical error propagation of Section~\ref{sec:analytic_error_propagation}.
  \href{https://github.com/sibirrer/glSNe/Notebooks/analytic_error_propagation.ipnb}{\faGithub}} %
\label{fig:number_forecast}
\end{figure*}

Figure~\ref{fig:number_forecast} shows the expected relative precision on $H_0$ as a function of the number of glSNe to be expected in the future for the three different micro-lensing scenarios and the two different external SNe samples considered in this work.
In about 2 years of the LSST survey when expecting $\sim 28$ glSNe, we forecast for the \textsc{realistic} scenario a $\sim 3\%$ precision on $H_0$.
With $\sim 150$ glSNe for the \textsc{Roman\_realistic} scenario we expect a 1\% precision on $H_0$. The precision of the external SNe sample substantially impacts the total error budget on $H_0$ for $>50$ glSNe in the \textsc{realistic} scenario.
These numbers in terms of years of LSST survey assume an optimal follow-up effort of the discovery candidates.

\section{Discussion} \label{sec:discussion}

The forecast results presented in Section~\ref{sec:forecasts} did only cover a limited range of possible systematics and opportunities regarding studying glSNe and measuring $H_0$. In this section, we discuss key systematics, other windows of opportunities, and we give some general recommendations driving the design requirements in future studies of glSNe to achieve a sub 2\% precision and accuracy of an $H_0$ measurement.

\subsection{Systematics}\label{sec:discussion_systematics}

\subsubsection{Selection effects}\label{sec:discussion_selection_effects}
Brightness selection effects in the discovery and follow-up analysis of glSNe systems may pose significant limitations in the standardizable magnification methodology. Bright glSNe are easier to discover and to follow-up. Such a selection can impact unlensed brightness selection as well as local lensing magnification selection.

In our forecast and methodology, we assume an identical unlensed peak SNe brightness distribution for the unlensed field sample and for the glSNe population ($\overline{m}_{\rm p}$). Unaccounted differences between the unlensed field sample, $\overline{m}_{\rm p, field}$, and the glSNe sample, $\overline{m}_{\rm p, glSNe}$, results in a differential shift in $H_0$ by
\begin{equation}
    \frac{\delta H_0}{H_0} = \frac{1}{2} \left(\overline{m}_{\rm p, glSNe} - \overline{m}_{\rm p, field} \right).
\end{equation}
Thus, an unaccounted relative selection effect of the field SNe and glSNE of 2\% results in a 1\% bias in $H_0$. Or in terms of an error budget, an uncertainty in the relative magnitude selection effect of 2\% results in an additional error term of 1\% on $H_0$ on top of the presented forecast results in Section~\ref{sec:forecasts}.

Local lensing magnification, a combination of micro-, milli-, and macro-lensing effects, may overall dominate the brightness selection. In particular, large (up-)scatter in brightness for rare micro-lensing events could significantly impact the selection function.
It is thus crucial to understand the micro-lensing selection effect.
Macro-lensing selection biases are less of an issue when performing the cosmographic analysis with time delays obtained by the identical selection function. However, when applying inferred mass profile constraints to lenses with different selection criteria, such as lensed quasars, the relative selection function comes into play.

\subsubsection{SNe dependence with redshift and host galaxies}
Beyond the glSNe systems and the required understanding of their selection function, breaking the MST and measuring $H_0$ also relies on an accurate and precise relative luminosity distance and intrinsic SNe distribution derived by an unlensed SNe data set. Such data sets are also used as a stand-alone cosmological probe or as a key component of a combined cosmological probe analysis, and their requirements and precision impact a glSNe+SNe analysis, as presented in this work.

For example, strong $\sim$0.1--0.2 magnitude dependence on the local
host-galaxy UV surface brightness, as reported by \cite{Rigault:2015}, needs to be understood when making inferences from high-redshift SNe Ia.
However, if there are reliable apparent magnitudes for unlensed field SNe available at the same redshifts as the glSNe, this can circumvent systematics limiting an SNe sample in measuring the late-time relative expansion history of the Universe.

We also note that with increased distance (higher redshifts) lensing effects also increasingly affect the apparent magnitudes of the field SNe sample as well. Relative selection effects (see Section~\ref{sec:discussion_selection_effects}) do also need to consider lensing selection effects in the field SNe sample.

We note that it is well known that the dust properties of SN~Ia hosts, parametrised by the total-to-selective absorption ratio, $R_V$, are very diverse and differ from the canonical value of the Milky Way of $R_V = 3.1$ \citep[e.g., see][for recent studies]{BS20, Thorp:2021, Johansson:2021}. Therefore, we require multiband data for each glSN in our sample to constrain the $R_V$ and color excesses in the host and lens galaxies. This is important since unresolved photometry alone has been shown to underestimate the inferred magnification, as seen for iPTF16geu \citep{Goobar:2017, Dhawan:2020}, mandating the need for optical and NIR coverage for each image of the glSNe.

\subsubsection{Gaussian uncertainty approximations}
In the forecast of this work, we assumed Gaussian uncertainties in the measurements (linear flux units), log-normal scatter in the intrinsic SNe peak brightness distribution, as well as Gaussian scatter in the milli- and micro-lensing magnifications.
The tails of the distributions need to be accurately captured to guarantee an unbiased joint inference\footnote{See e.g. Section 4.4 of \cite{Park:2021} about a discussion on tails in the external convergence distributions impacting combined constraints on $H_0$ for 200 quasar lenses.}.
In the current forecast, we explicitly distinguish between logarithmic and linear units and Gaussian likelihoods in either magnitude or flux units. This is not meant to be accurate for any specific scenario but primarily to emphasize the importance of accurately describing a likelihood or a posterior product.
Further care and emphasis must be undertaken in describing the probability density function (PDF) of the different components of the lensing magnifications. Specifically, non-Gaussian tails in the distributions, when combining a large set of glSNe, may significantly impact the resulting posterior PDF.
The hierarchical sampling and marginalization over population distributions further poses challenges in the accuracy of the likelihood evaluation and computational requirements. Gaussian or multivariate Gaussian distributions have the advantage of analytic solutions for marginalizations and likelihood evaluations, but the assumptions of Gaussian PDF's need to be tested to the requirements of the combined posterior densities.

\subsection{Opportunities}\label{sec:discussion_opportunities}
Aside from additional potential systematics considerations, there are also opportunities and circumstances that might increase the resulting precision on $H_0$ from glSNe relative to our fiducial forecast scenario. This section lists and briefly discusses a few of those opportunities.

\subsubsection{glSNe without a time delay}

The expected number of glSNe derived by \cite{Huber:2019} that we adopt in our forecast is, in part, based on the requirement to achieve a time-delay measurement. There are potentially many more glSNe expected to be discovered \citep[see e.g.][]{Goldstein:2019} where a precise time-delay measurement might not be expected. However, the availability of measured time delays is not the dominant source of uncertainty in our forecast. The primary information requirement to improve constraints on $H_0$ is foremost a precise absolute magnification measurement. 

\subsubsection{Galaxy--SNe lensing}
There is also a set of ``semi-strongly'' lensed SNe expected with a single magnified image available that is lensing through the outskirts of a lensing galaxy. An absolute magnification measurement remains possible in the absence of multiple images and such an enhanced sample might provide significant information of the more extended galaxy density profile and thus also constraining the physically plausible MST components \citep[see e.g.,][for such an analysis with a singly-lensed SNe in a cluster environment]{Rodney:2015}.
Such a probe is conceptually similar to galaxy--galaxy lensing and can possibly enhance the signal-to-noise in the very inner-most scales of galaxies where galaxy-shape information is less accessible and non-linear perturbations may arise on the distortion of the shapes \citep[see e.g.,][for work using magnifications of galaxies for such type of analysis]{Coupon:2013}.

\subsubsection{Other type of standardizable sources}
Our forecast has focused on SNe~Ia, in terms of the expected numbers, intrinsic scatter and light-curve properties to measure a peak brightness and a time delay. There are other transient sources that can be standardizable.
Different studies succeeded in constructing a Type~II SN Hubble diagram with a dispersion of $\sim10-14\%$ in distance \citep[e.g.,][]{Nugent:2006, Poznanski:2010, deJaeger:2015}. The more abundant Type II SNe may provide a valuable addition. Though the light curves of Type~II SNe are not as suited for time-delay measurements as with SNe~Ia, there might be advantages in measuring an absolute magnification effect with Type II SNe.

Beyond SNe, there are also gravitational waves (GW) that can be standardized remarkably well and thus may open-up opportunities beyond the capabilities of SNe. Repeated fast radio bursts (FRB's) may also provide the possibility for a standardization.
For GW and FRB's, one challenge will be the required astrometric precision to precisely determine the Fermat potential and macro-model magnification \citep[see e.g.,][]{BirrerTreu:2019}.

\subsubsection{Constraints from stellar kinematics}
In our forecast, we left out anticipated constraints from stellar dynamics measurements on density profiles and breaking the MST. In part because there is a larger literature on stellar kinematics in breaking the MST and existing recent literature providing a forecast for this methodology for the decade to come \citep{BirrerTreu:2021}. Another reason is to assess a kinematic-independent methodology in breaking the MST and thus constraints on the MST can be combined, provided both kinematics and standardizable magnifications are consistent.
We highlight that stellar kinematic measurements can be performed on the glSNe lenses once the glSNe have faded away and thus might provide similar, but independent, constraining power per glSNe.
Given that both methodologies are expected to provide about 1.5\% precision on $H_0$ in the next decade, this can result in stringent consistency checks, mitigation of currently non-anticipated systematics effects and establish a precise direct distance anchor of the Universe.

\subsection{Recommendations}\label{sec:disussion_recommendations}

Based on our forecast and the discussion of possible systematics and opportunities, we provide here some recommendations for the community to help guiding successful future strategies in providing both accurate and precise measurements of $H_0$ with glSNe. We focus on some aspects that either emerged directly from this work or deserves special emphasis.

\begin{enumerate}
    \item \textit{Perform follow-up observations for standardizable glSNe candidates regardless of the expected time-delay precision.}
    The precision on the mass profiles and hence $H_0$ relies on the ability of standardizable magnifications. Among the glSNe~Ia discoveries, those systems with low expected micro-lensing events are the most valuable in breaking the MST. A significant number of glSNe~Ia where at least one image is at lower magnification and lower projected stellar density are necessary, regardless of the time-delay precision
    \citep[see also][]{Foxley-Marrable:2018}. It is thus important to allocate significant follow-up efforts for those glSNe to be able to perform the analysis as forecasted in this work.

    \item \textit{Integrate weak and strong lensing SNe analysis.} To some extent, the division of the field SNe sample and the glSNe sample is an artificial cut in an underlying population of SNe that get lensed. Most lensing is weak with few percent magnification while the tails in the lensing magnification are effectively leading to glSNe. It is important to characterize the lensing effects across the entire spectrum to accurately describe the relative selection effects. With a more distant SNe sample, lensing effects may inevitably become more prominent also for the field SNe sample. 
    
    \item \textit{glSNe discovery strategy must provide a reproducible selection function.} Relative selection effects are possibly a dominant source of uncertainty or unaccounted systematics. Making use of the standardizable magnification effect to break the MST, it is crucial to understand and reproduce the relative selection effect to the percent level. A survey and discovery strategy must account for the feasibility to reproduce the selection function it contains. Known selection effects can then be mitigated by e.g. large-scale simulations \citep[see e.g.,][for the use for field SNe samples]{Scolnic:2016, Kessler:2017}.
    
    \item \textit{Extension of the hierarchical analysis to incorporate the astrophysics of micro-lensing.}
    The microlensing event statistics is by itself a phenomena that can probe the compact matter composition and fraction\citep[e.g.,][]{Schechter:2002, Kochanek:2004}. Correlations between stellar surface brightness and (microlensing) magnification events allows one to distinguish and measure the stellar initial mass function (IMF) and other forms of compact objects, such as primordial black holes (PBH).
    
\end{enumerate}

\section{Conclusions} \label{sec:conclusion}

Strongly lensed supernovae (glSNe) can provide, in addition to measurable time delays, lensing magnification constraints when knowledge about the unlensed apparent brightness of the explosion is imposed. In this paper, we discussed the theoretical aspects that allow absolute lensing magnifications to constrain a key property of the lensing mass profile that is insufficiently constraint with lensing-only data due to the mass-sheet degeneracy.
We then presented a hierarchical Bayesian analysis framework to combine a data set of SNe that are not strongly lensed and a data set of strongly lensed SNe with measured relative time delays. We jointly constrain (i) the unlensed apparent magnitude distribution of the population of SNe, (ii) the lens model profiles with the magnification ratio of lensed and unlensed fluxes on the population level, (iii) the relative expansion history of the Universe with the relative brightness of SNe with redshift, and (iv) $H_0$ with the time delays as an absolute distance indicator. 

We applied our joint inference framework on a future expected data set of glSNe from 10 years of the Rubin Observatory LSST in combination with a future unlensed SNe sample from the \textit{Roman Space Telescope}. We forecast that a sample of 144 glSNe with well measured time series and imaging data have the statistical power to measure $H_0$ to 1.0\% in the next decade. 

We discuss further expected covariant systematic uncertainties due to relative selection effects, dust extinction, and SNe redshift evolution.
We discussed strategies to mitigate systematics associated with using absolute flux measurements of glSNe to constrain the mass density profiles. Among the key systematic effect are relative selection biases in the discovery and usage of the glSNe and the unlensed SNe population due to micro-lensing magnification effects. We emphasize that for a 1\% precision on $H_0$, a 2\% overall accuracy in the standardization of apparent brightness distributions between SNe population in the field and the glSNe population needs to be achieved.
With an additional $1\%$ systematic uncertainty we forecast an overall precision on $H_0$ of $1.5\%$.

The methodology presented in this work is implemented in the public software \textsc{hierArc} and compatible with the hierarchical analysis by \cite{Birrer:2020}. The implementation allows one to combine different observational constraints self-consistently and can be adopted to the uncertain predictions of the expected glSNe depending on survey and follow-up strategies.

Using SNe is a promising and complementary alternative to using stellar kinematics observations to constrain the radial mass density profiles of strong lensing deflectors and can achieve comparable precision with independent assumptions and systematics.
Future surveys, such as the Rubin and \textit{Roman} observatories, will be able to discover the necessary number of glSNe, and with dedicated additional follow-up observations this methodology will provide precise constraints on mass density profiles and $H_0$. These constraints will be key to understand the source of the current Hubble tension, and will additionally provide insights into the formation and evolution of massive elliptical galaxies.

\begin{acknowledgments}
We thank Ariel Goobar, Justin Pierel, and Sherry Suyu for useful comments on an eariler version of the manuscript.
This research was supported by the U.S. Department of Energy (DOE) Office of Science Distinguished Scientist Fellow Program.
\end{acknowledgments}

%

\vspace{5mm}


\software{\textsc{lenstronomy} \citep{lenstronomy1, lenstronomy2},
          \textsc{hierArc} \citep{Birrer:2020},
          \textsc{astropy} \citep{astropy:2013, astropy:2018},  
          \textsc{emcee} \citep{emcee}.
          }

\section*{Data Availability}
The formalism and inference schemes presented in this work are implemented in \textsc{hierArc}\footnote{\url{https://github.com/sibirrer/hierArc}} and the scripts to reproduce the presented work is publicly available\footnote{\url{https://github.com/sibirrer/glSNe}}. Lensing calculations are performed with \textsc{lenstronomy}\footnote{\url{https://github.com/sibirrer/lenstronomy}}.



\appendix

\section{Lens model posteriors}\label{app:imaging_posteriors}
\begin{figure*}
  \centering
  \includegraphics[angle=0, width=100mm]{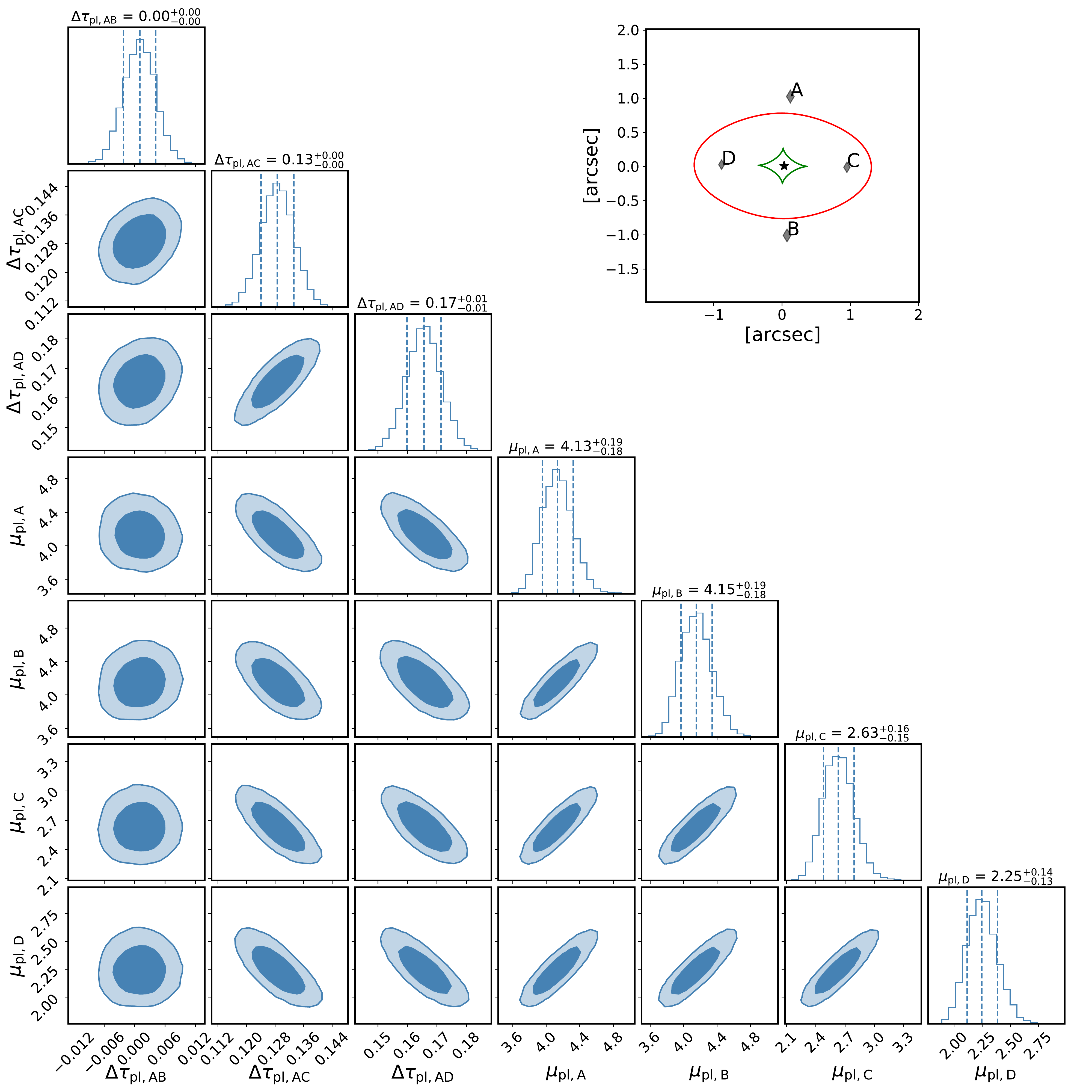}
  \caption{Mock image modeling posterior on the relative Fermat potential and lensing magnification between the image positions of a glSNe when fit by a PEMD+shear lens model for the cross configuration. The lens model parameters and uncertainties are presented in Table~\ref{table:lens_param}. The configuration of the image position (diamonds), inner caustic (green) and critical curve (red) are illustrated in the top right figure. \href{\notebooklink}{\faGithub}} %
\label{fig:cross_imaging_posteriors}
\end{figure*}

\begin{figure*}
  \centering
  \includegraphics[angle=0, width=100mm]{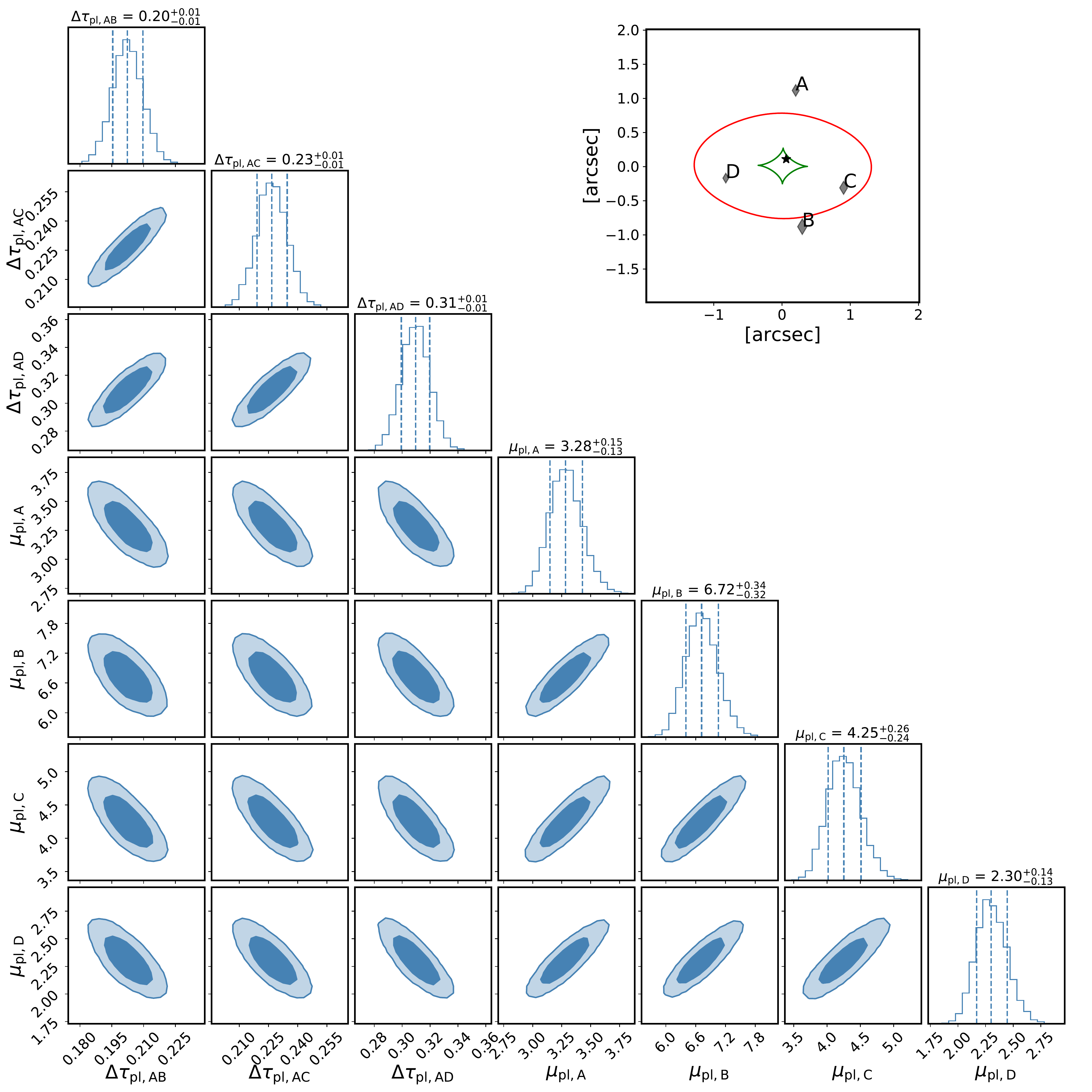}
  \caption{Mock image modeling posterior on the relative Fermat potential and lensing magnification between the image positions of a glSNe when fit by a PEMD+shear lens model for the fold configuration. The lens model parameters and uncertainties are presented in Table~\ref{table:lens_param}. The configuration of the image position (diamonds), inner caustic (green) and critical curve (red) are illustrated in the top right figure. \href{\notebooklink}{\faGithub}} %
\label{fig:fold_imaging_posteriors}
\end{figure*}

\begin{figure}
   \centering
  \includegraphics[width=.3\textwidth]{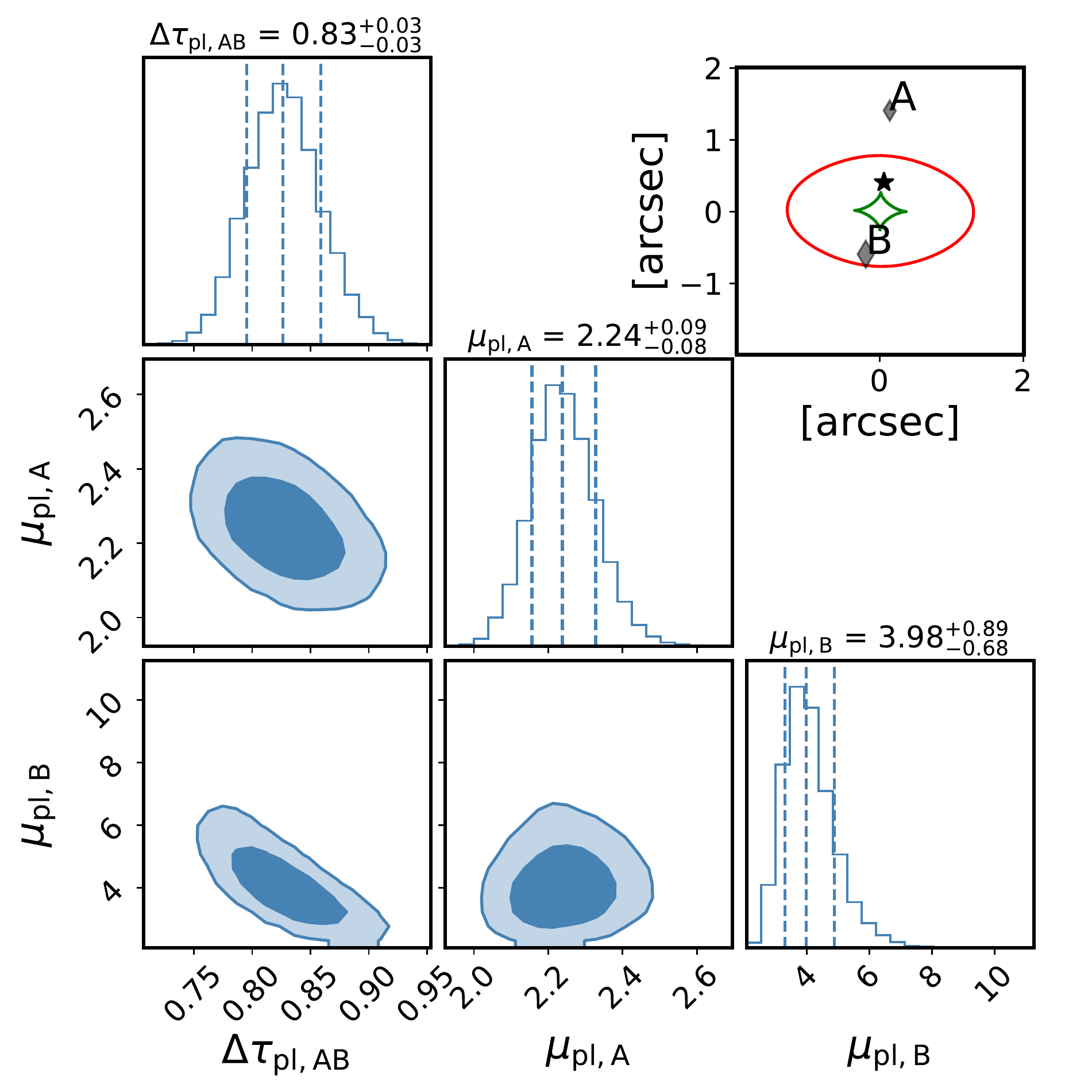}\hfill
  \includegraphics[width=.3\textwidth]{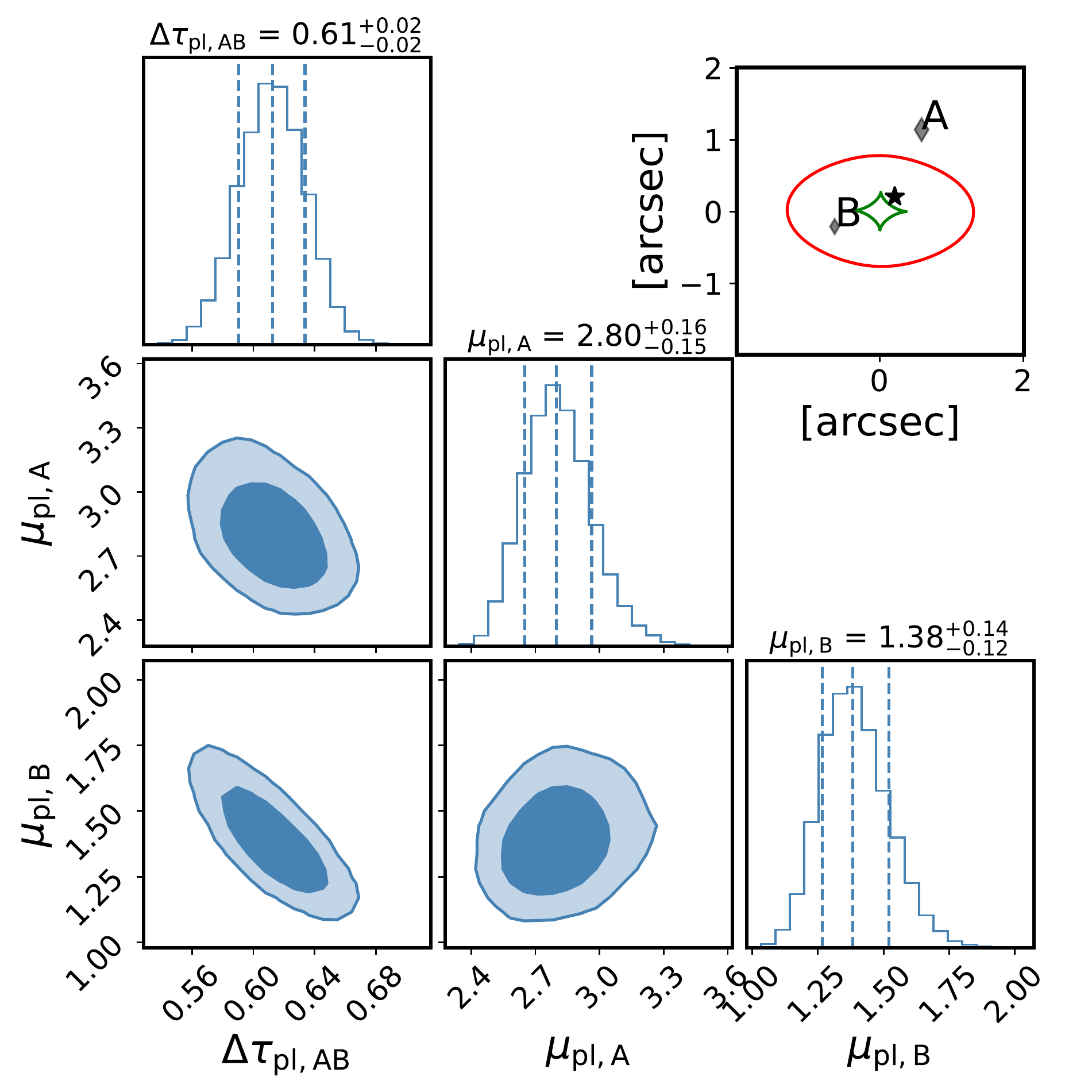}\hfill
  \includegraphics[width=.3\textwidth]{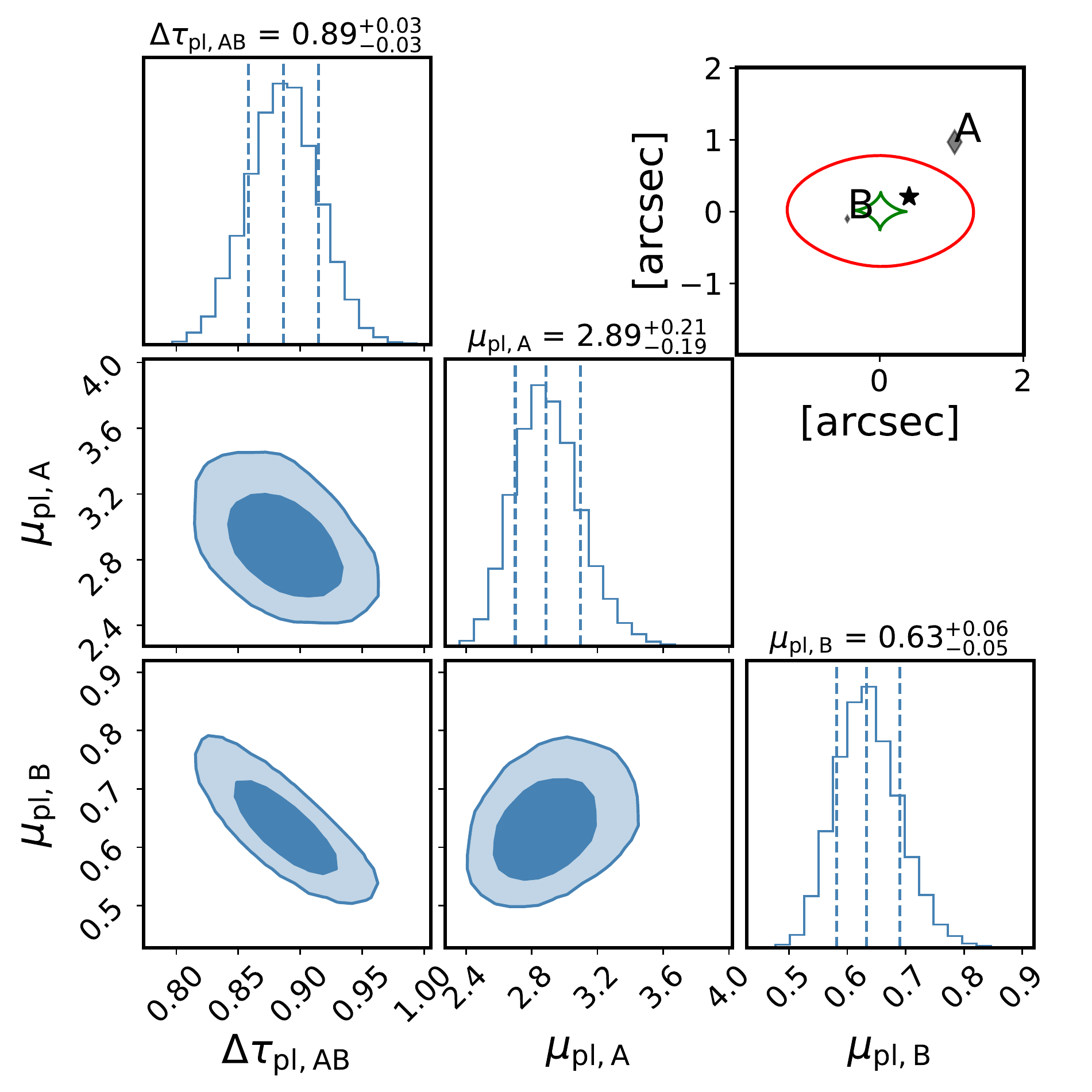}\hfill
  \caption{Mock image modeling posterior on the relative Fermat potential and lensing magnification between the image positions of a glSNe when fit by a PEMD+shear lens model for the double \#1 (left), \#2 (middle), and \#3 (right) configuration. The lens model parameters and uncertainties are presented in Table~\ref{table:lens_param}. The configuration of the image position (diamonds), inner caustic (green) and critical curve (red) are illustrated in the top right figure. \href{\notebooklink}{\faGithub}} %
\label{fig:double_imaging_posteriors}
\end{figure}

In this Appendix, we provide the posteriors of the Fermat potential differences and lensing magnification for the three quad and three double configuration lensing configuration based on the lens model and source position parameters and uncertainties of Table~\ref{table:lens_param}. We present the quadruply lensed configurations of the cross in Figure~\ref{fig:cross_imaging_posteriors}, and the fold configuration in  Figure~\ref{fig:fold_imaging_posteriors}. The cusp configuration is presented in the main body of the text in Figue~\ref{fig:cusp_imaging_posteriors}. The posteriors for the three double configurations are provided in Figure~\ref{fig:double_imaging_posteriors}.

\bibliography{glSNe}{}
\bibliographystyle{aasjournal}



\end{document}